\documentclass[10pt,onecolumn,oneside,notitlepage,final]{article}

\def \in #1 #2 {\int \limits_{#1}^{#2}}

\usepackage{amssymb}
\usepackage{amsmath}
\usepackage{graphicx} 
\usepackage{color}

\usepackage{amsmath}

\usepackage{pstricks}

\def\localinput#1{{
  \renewcommand{\documentclass}[2][dummy]{}
  \renewcommand{\usepackage}[2][dummy]{}
  \renewenvironment{document}{}{}
  \def\jobname{#1}
  \input{#1}
}}

\def\sla#1{\ooalign{\hfil\hspace{-0.1ex}\raise.2ex\hbox{$\not \phantom{#1}$}\hfil\crcr  $#1$}}

\def\siml{\hspace{0.3em}\raisebox{0.4ex}{$<$}\hspace{-0.75em}\raisebox{-.7ex}{$\sim$}\hspace{0.3em}}

\begin{document}

\begin{titlepage}

\date{}

\begin{flushright}
RUP-09-3
\end{flushright}
\vspace{12ex}
\begin{center}
 \huge{General Analysis of $B$ Meson Decay into Two Fermions}

\vspace{3ex}

\large{ Akihiro Matsuzaki\footnote{akihiro@rikkyo.ac.jp}
\\[4mm]
\small \textsl{ Department of Physics, Rikkyo University,}
 \\ \small \textit{Nishi-ikebukuro, Toshima-ku Tokyo, Japan, 171  }
 }

\vspace{3ex}


\end{center}

\begin{abstract}
    We study how to measure the current structure of the process that $B$ meson decays into two unstable fermions $\bar f_a$ and $f_b$ in model independent way.
    We use the momentum distributions of subsequent decay products affected by $\bar f_a f_b$ spin correlation.
    We have found the following:
 (1) We can extract the absolute values of two effective coupling constants from the opening angle between the particles decayed from $\bar f_a$ and $f_b$.
 (2) We can extract the real part of the interference from the energy distribution of one of the decayed particles from $\bar f_a$ or $f_b$.
 (3) No new information can be obtained from the energy distribution of two decayed particles from $\bar f_a$ and $f_b$.
 (4) The imaginary part of interference is extracted from the azimuthal angle asymmetry of final-state decay products.
 (5) If only one of two fermions is unstable, we can extract the real part of interference from each of the energy distribution and opening angle distribution.
   We show several simple examples.
\end{abstract}

\end{titlepage}

\section{Introduction}

    A huge number of $B$ mesons are produced in B-factories.
    They are used to confirm the Standard Model (SM).
    Almost of all the results suggest that the SM, and especially, Kobayashi-Mazkawa ansatz are reliable.
    Recently, we search for rare events and SM-forbidden phenomena in B-factories with high statistics.
    However, new physics has not been seen.

    To discover them, it is important to search through many modes and many physical quantities.
    They are, for instance, CP asymmetry, forward-backward asymmetry, left-right asymmetry, energy distribution, and angular distribution.
    We want to detect not only the decay width but also these quantities.
    Also, we want to analyze as many channels as possible using the unified form for simplicity, facility, and practicality.
    Another important thing to discover the new physics is making the reliable SM prediction especially for the non-perturbative QCD effect.
    Also for this purpose, determining many physical quantities is significant.

    In this paper, we consider the general $B\to \bar f_a f_b$ decay modes, where $f_a$ and $f_b$ are arbitrary fermions and $\bar f_a$ is the antiparticle of $f_a$.
    The CP violation can be measured in some of these modes \cite{Bto2baryons}.
    These modes can be divided in two types.
    One is the leptonic decay modes and another is the baryonic decay modes.
    The SM prediction in leptonic modes are \cite{PRD55.2768, Btotau} 
\begin{align} \begin{split}\label{eq1}
 Br( B_d^0 &\to \tau^+ \tau^-) \simeq 2.8\times10^{-8}
\\Br(B_s^0 &\to \tau^+ \tau^-) \simeq 8.9\times10^{-7}.
\end{split} \end{align}
    The experimental upper bound is \cite{PDG1}
\begin{align} \begin{split}
  Br(B_d^0 &\to \tau^+ \tau^-) <4.1\times10^{-3}.
\end{split} \end{align}
    On the other hand, the branching ratios of baryonic modes are predicted as \cite{ph/0512335}
\begin{align} \begin{split}
  Br(B_d^0 &\to \bar \Xi_c^- \Lambda_c^+) \sim 2.0\times10^{-3}
\\Br(B_u^+ &\to \bar \Xi_c^0 \Lambda_c^+) \sim 2.2\times10^{-3}.
\end{split} \end{align}
    The experimental upper bounds are for example, \cite{PDG1}
\begin{align} \begin{split}
\\Br(B_d^0 &\to \Delta^0 \bar \Lambda) < 9.3\times10^{-7}
\\Br(B_d^0 &\to \bar \Lambda \Lambda) <3.2 \times 10^{-7} 
\\Br(B_d^0 &\to \bar \Lambda_c^- p )=
    (2.1 \raisebox{0.9ex}{+0.7}\hspace{-2.05em}\raisebox{-0.95ex}{$-$0.5})\times10^{-5}
\\Br(B_d^0 &\to \bar \Lambda_c^- \Lambda_c^+) <6.2 \times 10^{-5}, 
\\
\\Br(B_u^+ &\to \Delta^+ \bar \Lambda) <8.2 \times 10^{-7} 
\\Br(B_u^+ &\to \bar\Xi_c^0 \Lambda_c^+)Br(\bar\Xi_c^0 \to \bar\Xi^+ \pi^-)  
=(5.6 \raisebox{0.9ex}{+2.7}\hspace{-2.05em}\raisebox{-0.95ex}{$-$2.4}) \times 10^{-5} 
\\Br(B_u^+ &\to \bar\Xi_c^0 \Lambda_c^+)Br(\bar\Xi_c^0 \to  K^+ \pi^-)  
=(4.0 \pm1.6) \times 10^{-5}. 
\end{split} \end{align}

    $B_d^0 \to \bar \Lambda_c^- p$ and $B_u^+ \to \bar\Xi_c^0 \Lambda_c^+$ have already seen.
    Also, some other modes are predicted to be seen in near future by the SM or other models.
    Comparing the experimental result with the model predictions of current structure, we try to discover new physics, select a reasonable model, and consider the non-perturbative QCD effects.


    The modes which decay into unstable particles decrease the efficiency since it is difficult to detect the events, however these modes have the advantage in correlation detection.
    The correlation is detected as momentum distribution of $a$ and $b$, which are the decay products of $\bar f_a$ and $f_b$, respectively.
    When we deal with these modes, we have to consider the whole process of
\begin{align} \begin{split}
B \to &\bar f_a + f_b \to b+\mathrm{anything} \\
&\hspace{0.3em}{}^\lfloor \hspace{-0.5em} \longrightarrow a+\mathrm{anything}
\end{split} \end{align}
    because we cannot detect the intermediate state $\bar f_a + f_b$.

\subsection{$B_q \to \bar f_a f_b$ decay}
    From the partially conserved axial current relation, the general $B_q \to \bar f_a f_b$ decay amplitude is given by \cite{PRD55.2768}
\begin{align} \begin{split}
A_q=if_B m_B G_F [(C_P^q+\frac{m_b+m_a}{m_B}C_A^q)(\bar f_b \gamma_5 f_a)
+(C_S^q+\frac{m_b-m_a}{m_B}C_V^q)(\bar f_b f_a)],
\end{split} \end{align}
    where $f_B$, $m_B$, and $G_F$ are $B$ meson decay constant, $B$ meson mass, and the Fermi constant, respectively;
    $m_a$ and $m_b$ are $\bar f_a$ and $f_b$ masses, respectively;
    $C_P^q$, $C_S^q$, $C_A^q$, and $C_V^q$ are the complex coefficients of pseudo scalar, scalar, axial, and vector currents, respectively;
    The superscript $q$ represents the valence $u$, $d$, $s$ or $c$ quark in $B$ meson.     

    In charged $B$ meson decays, we simply set
\begin{align} \begin{split}\label{Cs}
C_1 &\equiv  C_P^q  +\frac{m_b+m_a}{m_B} C_A^q  \\
C_2 &\equiv  C_S^q  +\frac{m_b-m_a}{m_B} C_V^q.
\end{split} \end{align}     
    On the other hand, in neutral $B$ meson decays, considering the $B^0 - \bar B^0$ mixing effect, we set \cite{ph/0511079}-\cite{BS}
\begin{align} \begin{split}
|B^0(t)\rangle&=g_+(t)|B^0\rangle+\frac{q}{p}g_-(t)|\bar B^0\rangle,
\\g_\pm(t)&=\frac{1}{2}e^{-im_B t}e^{-\frac{1}{2}\Gamma_B t}\Bigl[1\pm e^{-i\Delta m_B t}e^{\frac{1}{2}\Delta\Gamma_B t}\Bigr],
\end{split} \end{align}
    where; $t$ is the time started when $B^0$ is created;
    $q/p$ is the ratio of $\bar B^0$ to $ B^0$ in $B^0$ mass eigenstate; 
    $\Gamma_B$ is the $B^0$ total decay width; $\Delta m_B$ and $\Delta\Gamma_B$ are the mass deference and decay width difference between heavier and lighter $B^0$ mesons;
    Hence, the time dependent effective amplitude takes the form
\begin{align} \begin{split}
 A_q(t)=if_B m_B G_F \Bigl[\widetilde{C}_1(\bar f_b \gamma_5 f_a) +\widetilde{C}_2(\bar f_b f_a)\Bigr],
\end{split} \end{align}
    where
\begin{align} \begin{split}\label{td}
\widetilde{C}_1&\equiv\Bigl\{g_+(t)C_1+\frac{q}{p}g_-(t)\bar C_1\Bigr\},
 \\\widetilde{C}_2&\equiv\Bigl\{g_+(t)C_2+\frac{q}{p}g_-(t) \bar C_2\Bigr\}.
\end{split} \end{align}
%
%
%
%
%
    These parameters appear in the differential decay width in the form of $|\widetilde{C}_1|^2$, $|\widetilde{C}_2|^2$, $Re[\widetilde{C}_1\widetilde{C}_2^*]$, and $Im[\widetilde{C}_1\widetilde{C}_2^*]$. 
    These quantities depend on the decay time.
    However, for $|p/q|=|C_1/\bar C_1|=|C_2/\bar C_2|=1$ and $\Delta \Gamma_B=0$, integrating over the time and summing over $B^0$ decays and $\bar B^0$ decays, these quantities becomes $2|C_1|^2/\Gamma_B$, $2|C_2|^2/\Gamma_B$, $(Re[C_1C_2^*]+Re[\bar C_1 \bar C_2^*])/\Gamma_B$, and $(Im[C_1C_2^*]+Im[\bar C_1\bar C_2^*])/\Gamma_B$, respectively.    
%
%
    Hence, we omit the time dependence and the tildes on $C_1$ and $C_2$ in most of the rest of this paper.

    We want to give the $B\to \bar f_a f_b$ partial decay width in which $\bar f_a$ and $f_b$ have particular polarizations.
    Thus, we introduce the polarization vectors $s^a$ and $s^b$ of $\bar f_a$ and $f_b$, respectively.
    These vectors have the constraints $(s^a)^2=(s^b)^2=-1$ and $s^a\cdot k_{f_a}=s^b\cdot k_{f_b}=0$, where $k_{f_a}$ and $k_{f_b}$ are $\bar f_a$ and $f_b$ momenta, respectively.

    In $B$ rest frame, the differential decay width of $B \to \bar f_a(s^a) f_b(s^b)$ is given by
\begin{align} \begin{split}
\frac{d\Gamma}{ d\Omega }  
=&\frac{f_B^2G_F^2}{32\pi^2}|\mathbf{p}|
\Bigl\{D_1+D_2( s_x^a s_x^b + s_y^a s_y^b )+D_3 s_z^as_z^b
\\& \hspace{4.5em}+D_4(\frac{m_b}{m_a} s_z^a -\frac{m_a}{m_b} s_z^b )+D_5( s_x^a s_y^b - s_x^b  s_y^a )\Bigr\}, 
\end{split} \end{align}
    where
\begin{align} \begin{split}\label{Ds}
 |\mathbf{p}|&=\frac{\sqrt{(m_B^2-(m_a-m_b)^2)(m_B^2-(m_a+m_b)^2)}}{2m_B},
\\D_1&=        |C_1|^2\frac{m_B^2-(m_a-m_b)^2}{2} +|C_2|^2\frac{m_B^2-(m_a+m_b)^2}{2},  
\\D_2&=       -|C_1|^2\frac{m_B^2-(m_a-m_b)^2}{2} +|C_2|^2\frac{m_B^2-(m_a+m_b)^2}{2},
\\D_3&=       -D_1,
\\D_4&=-2 Re[C_1C_2^*] m_a m_b \gamma_a\gamma_b(\beta_a+\beta_b),
\\D_5&=-2 Im[C_1C_2^*] m_a m_b \gamma_a\gamma_b(\beta_a+\beta_b),
\end{split} \end{align}
\begin{align} \begin{split}
\beta_a\equiv\frac{|\mathbf{k}_{f_a}|}{k_{f_a}^0},
&\hspace{3em}\gamma_a\equiv\frac{1}{\sqrt{1-\beta_a^2}}=\frac{k_{f_a}^0}{m_a},
\\\beta_b\equiv\frac{|\mathbf{k}_{f_b}|}{k_{f_b}^0},
&\hspace{3em}\gamma_b\equiv\frac{1}{\sqrt{1-\beta_b^2}}=\frac{k_{f_b}^0}{m_b},
\end{split} \end{align}
    and $\Omega$ is the solid angle of $k_{f_a}$.

    The general $B\to \bar f_a f_b \to a+b+\mathrm{anything}$ differential decay width is written as 
\begin{align}\begin{split}  
\frac{d\Gamma}{ d\Omega \   d^3 k_a  \ d^3 k_b }  
=& \hspace{ 0.5em}
\raisebox{-1ex}{\text{{\huge \textsf{S}}}}\hspace{-1.5em}   \raisebox{-3ex}{$\scriptstyle {s^a , s^b}$}     \hspace{ 0.5em}
 \sum _{\pm s^a ,  \pm s^b}\frac{d\Gamma\bigl(B \to \bar f_a(s^a) f_b(s^b)\bigr)}{dt\ d\Omega} \\ &\hspace{4em}\times
\frac{dBr\bigl( \bar f_a(s^a)  \to a+\mathrm{anything}\bigr)}{d^3 k_a}
 \\ &\hspace{4em} \times
\frac{dBr\bigl(f_b(s^b)        \to b+\mathrm{anything}\bigr)}{d^3 k_b},
\end{split} \end{align}
    where \textsf{S} implies sum over polarizations. 
$k_a$ and $k_b$ are the momenta of the particle $a$ and $b$ in $\bar f_a$ and $f_b$ rest frame, respectively.
    The differential branching ratios of $\bar f_a$ and $f_b$ are written in Appendix \ref{App:f-decay}. 

\begin{figure}[ht]
  \begin{center}
    \includegraphics[keepaspectratio=true,height=50mm]{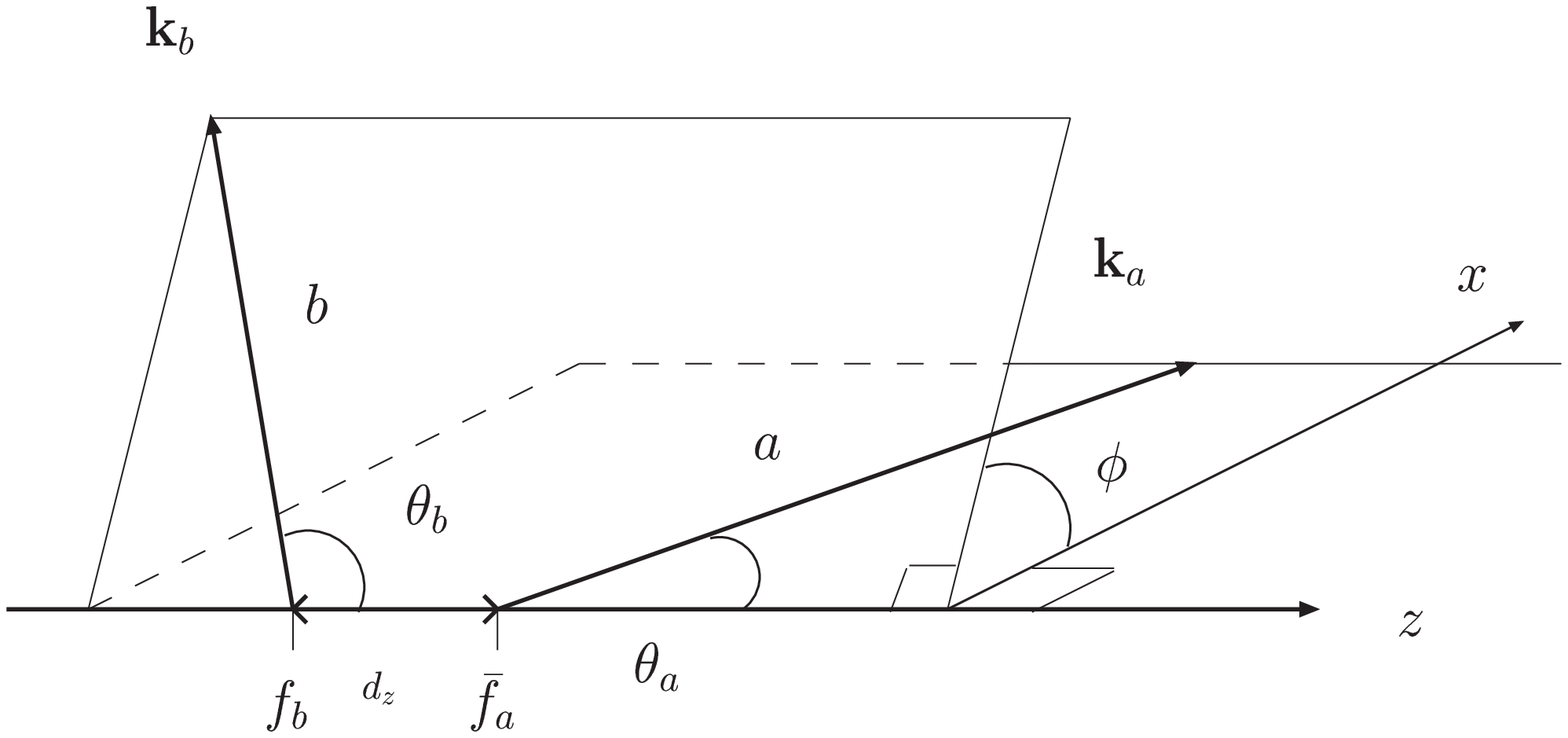}
  \end{center}
  \caption{$B$ meson rest frame.
  $\bar f_a$ is oriented in the positive $z$-axis direction.
  The zenith angles of $a$ and $b$ directions are $\theta_a$ and $\theta_b$, respectively.
  The azimuthal angle between $a$ and $b$ is $\phi$.
  The distance between $\bar f_a$ and $f_b$ decay points is $d_z$.
  $a$ and $b$ have the momenta $\mathbf{k}_a$ and $\mathbf{k}_b$, respectively.
}
  \label{Frame-3}
\end{figure}

    In writing the explicit form of the decay width, we will use the following notation (See Fig. \ref{Frame-3}.):
    In $B$ rest frame, $\bar f_a$ is oriented in the positive z-axis direction. 
    The zenith angles of $a$ and $b$ directions in $B$ rest frame are $\theta_a$ and $\theta_b$, respectively.
    The azimuthal angle between $a$ and $b$ directions is $\phi$.
    $d_z$ is the distance between $\bar f_a$ and $f_b$ decay points.

    In the massless limit of $a$ and $b$, we obtain the general formula   
\begin{align}\begin{split}\label{master-formula}  
&\frac{d\Gamma}{  dy_a d\Omega_a \ dy_b d\Omega_b} 
\\ = &   
Br_a\frac{y_a^2}{4\pi  \lambda_a}
Br_b\frac{y_b^2}{4\pi  \lambda_a}
 \frac{f_B^2G_F^2}{2\pi}|\mathbf{p}|
\\&\times\biggl[D_1  G_1^a(y_a)  G_1^b(y_b)
+D_4\Bigl\{\frac{m_b}{m_a} \cos\theta_a G_1^b(y_b)G_2^a(y_a) +\frac{m_a}{m_b}  \cos\theta_b G_1^a(y_a)G_2^b(y_b)\Bigr\}
\\&\hspace{2em}-\bigl(D_2 \sin\theta_a \sin\theta_b \cos\phi
-D_1 \cos\theta_a \cos\theta_b 
+D_5 \sin\theta_a \sin\theta_b \sin\phi \bigr) G_2^a(y_a)G_2^b(y_b)\biggr],            
\end{split} \end{align}
    where
\begin{align}\begin{split}  \label{yayb}
   Br_a &= Br(\bar f_a \to a+\mathrm{anything}),
\\ Br_b &= Br( f_b \to b+\mathrm{anything}),
\\y_a&=\frac{2E_a}{m_a},\ \ \ y_b =\frac{2E_b}{m_b},
\end{split} \end{align}
    where $E_a$ and $E_b$ are $a$ and $b$ energy in $\bar f_a$ and $f_b$ rest frames, respectively;
    $G_{1,2}^{a,b}(y_{a,b})$ are the functions which are defined in Appendix \ref{App:f-decay}.
    The massless condition of particles $a$ and $b$ are reasonable because most of $\tau$ decay into $\mu$, $e$, or pions, and substantial unstable baryons decay into a lighter baryon and pions, photons, and/or leptons.
    They have at most about 100 MeV masses, which are enough smaller than the masses of $\tau$ and any baryons.

 
    Using the general formula (\ref{master-formula}), we first derive the partial decay width.
    Integrating over $dy_a d\Omega_a dy_b d\Omega_b$, we have
\begin{align}\begin{split}  \label{total}
\Gamma=Br_a Br_b \frac{f_B^2G_F^2}{2\pi}|\mathbf{p}|D_1.
\end{split} \end{align}

    This width contains the factor $D_1$.
    We determine this coefficient, first.
    However, we want to know the relation between $|C_1|$ and $|C_2|$.
    Moreover, we want to know how is the relative phase between $C_1$ and $C_2$.
    That is what we will do in this paper.

    This paper is organized as follows:
    In Section 2, we consider the energy distribution of $a$ to determine $Re[C_1C_2^*]$. 
    In Section 3, we consider the distribution of opening angle between $a$ and $b$ to determine $|C_1|$ and $|C_2|$, separately. 
    In Section 4, we consider the azimuthal angle asymmetry of $a$ and $b$ to determine $Im[C_1C_2^*]$. 
    In Section 5, we discuss the case that $f_b$ is a stable fermion.
    In Section 6, we show some examples of baryonic mode.
    In Section 7, we summarize our analysis.

\section{Energy Distribution}\label{Energy Distribution}

    In this section, we study the energy distribution of the final-state particle $a$ or $b$.
    For definiteness, let's say that we want to investigate the $a$ energy distribution.

    The prescription to derive the energy distribution formula in $B$ rest frame is as follows \cite{sanda}:
    First, we multiply the delta function $\delta \bigl( x_a-y_a(1+\beta_a\cos\theta_a)/2\bigr)$ by Eq. (\ref{master-formula}), where $x_a=E_a'/E_{f_a}$, and $E_{f_a}$ and $E_a'$ are $\bar f_a$ and $a$ energy in $B$ rest frame, respectively.
    $x_a$ means a normalyzed energy of particle $a$ in $B$ rest frame.
    Next, we integrate over $dy_a d\Omega_a dy_b d\Omega_b$.
    Then, we have
\begin{align}\begin{split} \label{E-D} 
\frac{1}{\Gamma}\frac{d\Gamma}
{ d x_a   } 
=\int d y_a  
\frac{1}{\beta_a \lambda_a}
\Bigl\{ y_a G_1^a(y_a)  
+\frac{D_4}{D_1}\frac{ m_b}{m_a \beta_a}( 2x_a-y_a)  G_2^a(y_a) 
\Bigr\}.            
\end{split} \end{align}
    Here, $\int d y_a$ means 
\begin{align} \begin{split}
 \int d y_a
 =\in \frac{2x_a}{1+\beta_a} \frac{2x_a}{1-\beta_a}  d y_a 
 \theta[x_a]\theta[\frac{1-\beta_a}{2}-x_a]
 +
  \in \frac{2x_a}{1+\beta_a} 1  d y_a 
 \theta[x_a-\frac{1-\beta_a}{2}]\theta[\frac{1+\beta_a}{2}-x_a].
\end{split} \end{align}
    The expression (\ref{E-D}) suggests that $a$ energy dependence can be used to determine the coefficient $D_4$, which contains $Re[C_1C_2^*]$.
    We note that $b$ energy dependence can also be used to determine $D_4$, similarly.
    However, no new information is obtained by the energy distributions of both of $a$ and $b$, namely, $d\Gamma/( dx_a dx_b)$. 
    This is because $D_2$ and $D_5$ terms in the general formula (\ref{master-formula}) vanish when we integrate over the azimuthal angle $\phi$.

\subsection{Example 1 - $\tau^+$ Decays into $\mu^+ \nu_\mu \bar\nu_\tau$}

    As a simple example, we calculate $\mu^+$ energy distribution of 
\begin{align} \begin{split}
B^0\to&\tau^+ +\tau^-
\\&\hspace{0.3em}{}^\lfloor \hspace{-0.5em} \longrightarrow 
\mu^++\nu_\mu+\bar\nu_\tau. 
\end{split} \end{align}
    In this case, we can set $G_1^a(y_a)=3-2y_\mu$, $G_2^a(y_a)=2y_\mu-1$, $\lambda_a=\lambda_b=\frac{1}{2}$, and $\beta_a=\beta_b= \sqrt{1-4m_\tau^2/m_B^2} \equiv \beta$.
    Hence, we have 
\begin{align}\begin{split}  \label{emu}
\frac{1}{\Gamma}&\frac{d\Gamma}
{ d x_\mu   }  \\
=& \frac{2 }{\beta } 
  \Biggl[
\biggl\{ 
\frac{8  \beta x_\mu^2 \left(9(1-\beta^2)-4(3+\beta^2) x_\mu\right)}{3 \left(1-\beta^2\right)^3}%
\\& \hspace{2em}  + \frac{2Re[\widetilde{C}_1\widetilde{C}_2^*]}{ |\widetilde{C}_1|^2 +|\widetilde{C}_2|^2 \beta^2} 
   \frac{8  \beta^3 x_\mu^2 \left(16 x_\mu-3(1-\beta^2)\right)}{3 \left(1-\beta^2\right)^3}
 \biggr\}
\theta[x_\mu]\theta[\frac{1-\beta}{2}-x_\mu]
\\& \hspace{1em}
+\biggl\{ 
 \frac{5(1+ \beta)^3-4 \bigl(9(1+ \beta)-8x_\mu\bigr) x_\mu^2}{6(1+\beta)^3}
\\& \hspace{3em}  + \frac{2Re[\widetilde{C}_1\widetilde{C}_2^*]}{ |\widetilde{C}_1|^2 +|\widetilde{C}_2|^2 \beta^2} 
   \frac{ \left((1+\beta)^3-12 (1+2 \beta)(1+\beta) x_\mu^2 +16 (1+3 \beta) x_\mu^3\right)}{6
   (1+\beta)^3}
 \biggr\}
\\& \hspace{1em} \times \theta[x_\mu-\frac{1-\beta}{2}]\theta[\frac{1+\beta}{2}-x_\mu]\Biggr].
\end{split} \end{align}
    Here, we put tildes on $C_1$ and $C_2$.
    Integrating $\Gamma$ and $d\Gamma/dx_\mu$ over the time, and summing them over $B^0$ decays and $\bar B^0$ decays, 
    the energy distribution is represented by Eq. (\ref{emu}) replacing $|\widetilde{C}_1|^2$, $|\widetilde{C}_2|^2$, and $Re[\widetilde{C}_1\widetilde{C}_2^*]$ with  $|{C}_1|^2$, $|{C}_2|^2$, and $(Re[{C}_1 {C}_2^*]+Re[\bar{C}_1 \bar{C}_2^*])/2$, respectively.

\begin{figure}[ht]
  \begin{center}
    \includegraphics[keepaspectratio=true,height=50mm]{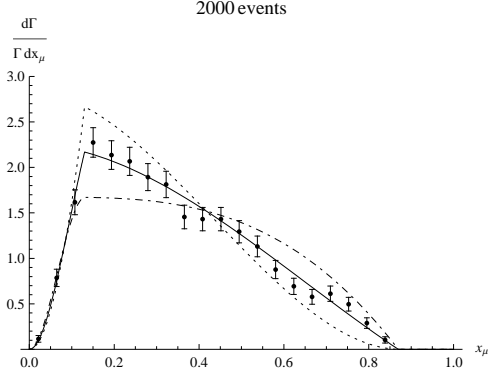}
  \end{center}
  \caption{The $\mu$ energy distribution of $B^0\to\tau^+ \tau^-$ and subsequently $\tau^+\to\mu^+(e^+)+\nu_{\mu(e)} +\bar \nu_\tau$ or $\tau^-\to\mu^-(e^-)+\bar\nu_{\mu(e)} +\nu_\tau$ decay, and their CP conjugate.
    The horizontal axis is the normalized $\mu$ energy $x_\mu$.
    The vertical axis is the time integrated differential decay width $d\Gamma/dx_\mu$ over the time integrated partial width $\Gamma$.  
    We set $|{C}_1|=|{C}_2|=1$.
    The solid line, dashed line, and dot-dashed line represent $(Re[C_1 C_2^*]+Re[\bar C_1 \bar C_2^*])/|2C_1C_2|=\{0,1,-1\}$ case, respectively.
    The MC result in a sample of 2000 events for $(Re[C_1 C_2^*]+Re[\bar C_1 \bar C_2^*])/|2C_1C_2|=0$ is $(Re[C_1 C_2^*]+Re[\bar C_1 \bar C_2^*])/|2C_1C_2|=  0.11\pm0.10$.        
    }
\label{1e-mu-c11-c21}
\end{figure}

\begin{figure}[ht]
  \begin{center}
    \includegraphics[keepaspectratio=true,height=50mm]{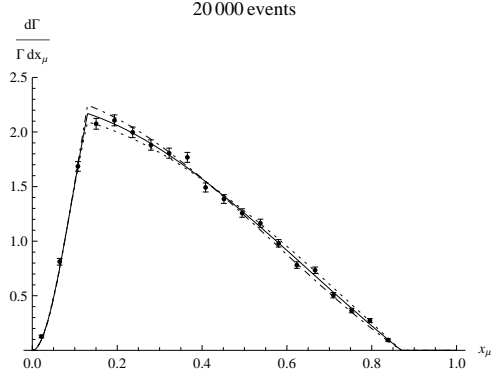}
  \end{center}
  \caption{The absolute values of coefficients are $|{C}_1|=1$ and $|{C}_2|=0.1$.
            The number of event is 20000. 
            Others are the same as Fig. \ref{1e-mu-c11-c21}.
            The MC result is $(Re[C_1 C_2^*]+Re[\bar C_1 \bar C_2^*])/|2C_1C_2|=-0.05\pm0.21$.
}
\label{1e-mu-c11-c20.1}
\end{figure}

\begin{figure}[ht]
  \begin{center}
    \includegraphics[keepaspectratio=true,height=50mm]{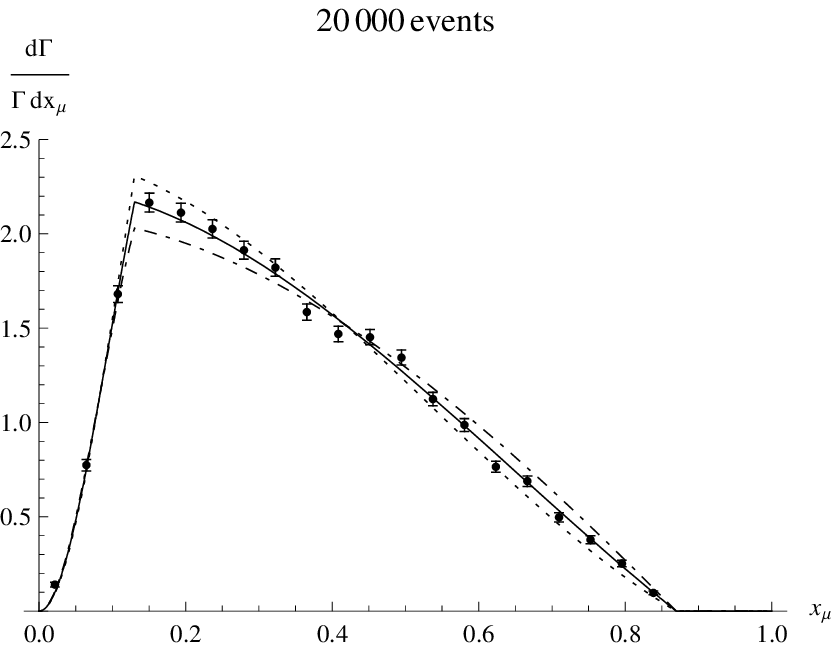}
  \end{center}
  \caption{The absolute values of coefficients are $|{C}_1|=0.1$ and $|{C}_2|=1$.
            The number of event is 20000. 
            Others are the same as Fig. \ref{1e-mu-c11-c21}.
            The MC result is $(Re[C_1 C_2^*]+Re[\bar C_1 \bar C_2^*)/|2C_1C_2|=0.17\pm0.12$.
}
\label{1e-mu-c10.1-c21}
\end{figure}

    We depict this distribution and perform a Monte Carlo simulation (MC) to estimate the error of $(Re[C_1 C_2^*]+Re[\bar C_1 \bar C_2^*])/|2C_1C_2|$ in Figs. \ref{1e-mu-c11-c21}-\ref{1e-mu-c10.1-c21}.

Fig. \ref{1e-mu-c11-c21} represents the $|{C}_1|=|{C}_2|=1$ case.
    Similarly, Fig. \ref{1e-mu-c11-c20.1} and Fig. \ref{1e-mu-c10.1-c21} represent the $\{|{C}_1|,|{C}_2|\}=\{1,0.1\}$ and $\{|{C}_1|,|{C}_2|\}=\{0.1,1\}$ cases, respectively.
    The interference effect emerges when $|{C}_1|\simeq|{C}_2|\beta$.
    We note here that in these figures, $Re[{C}_1 {C}_2^*]$ independent point where $x_\mu\simeq 0.4$ is one of the solutions of identity,
$(1 + \beta)^3 - 12 (1 + 2 \beta) (1 + \beta) x_\mu^2 + 16 (1 + 3 \beta) x_\mu^3 = 0$.

    In Fig. \ref{1e-mu-c11-c21}, the MC is performed in a sample of 2000 events for $(Re[C_1 C_2^*]+Re[\bar C_1 \bar C_2^*])/|2C_1C_2|=0$.
    The number of events are given as follows:
    The Super KEKB will make about $50$ ab${}^{-1}$ integrated luminosity.
    The $e^+ +e^-\to\Upsilon(4S) \to B_d^0 \bar B_d^0 $ cross section is about $10^{-33}$ cm${}^2$. 
    The $ B_d^0 \to \tau^+\tau^- $ branching ratio is in Eq. (\ref{eq1}).
    The $ \tau^+ \to \mu^+ \nu_\mu \bar \nu_\tau$ branching ratio and the $ \tau^+ \to e^+ \nu_e \bar \nu_\tau$ branching ratio are about $0.174$ and $0.178$.
    These are essentially the same events for the massless limit of daughter fermions.
    Therefore, about 2000 events will be available. 
    The efficiency of this mode is in fact very low.
    However, we here just ignore it. 
    The MC result is $(Re[C_1 C_2^*]+Re[\bar C_1 \bar C_2^*])/|2C_1C_2|=0.11\pm0.10$.  
 
    In Figs. \ref{1e-mu-c11-c20.1} and \ref{1e-mu-c10.1-c21}, the MC is performed in a sample of 20000 events for $(Re[C_1 C_2^*]+Re[\bar C_1 \bar C_2^*])/|2C_1C_2|=0$.
    The MC results in $(Re[C_1 C_2^*]+Re[\bar C_1 \bar C_2^*]) /$ $ |2C_1C_2|$ are $-0.05\pm0.21$ and $0.17\pm0.12$, respectively.

    In the SM, $|C_1|\gg |C_2|$ \cite{PRD55.2768}.
    Then, the coefficient $2Re[\tilde C_1 \tilde C_2^*]/(|\tilde C_1|^2+\beta|\tilde C_2|^2)$ is nearly zero.
    This situation is also realized if we set $Re[ C_1  C_2^*]=Re[\bar C_1 \bar C_2^*]=0$.
    Therefor, the shape of distribution is the same as $(Re[C_1 C_2^*]+Re[\bar C_1 \bar C_2^*)/|2C_1C_2|=0$ case.



    $B\to\tau^+\tau^-$ energy distribution is very interesting since we can investigate the current structure of new physics. 

    For instance, Ref. \cite{arXiv:0909.0863} expresses the Higgs induced operators for the transition $b\to s \mu^+\mu^-$.
    It is easy to transform them for the transition $b\to d \tau^+\tau^-$.
    Concletely, it is realized by the deformations $m_\mu \to m_\tau$, $F_{23} \to F_{13}$, $F_{32}^* \to F_{31}^*$, $ \mu \to \tau$, and $s \to d$.
    After that, the coefficients $C_1$ and $C_2$ in this paper are given by 
\begin{align} \begin{split}\label{eq 0909.0863}
C_1  &=-\frac{1}{2G_F}\frac{m_\tau}{2v \cos\beta}\frac{\sin\beta}{M_A^2}(F_{13}+F_{31}^*)\\
C_2  &=-\frac{1}{2G_F}\frac{m_\tau}{2v \cos\beta}\left(\frac{\sin(\alpha-\beta)\cos\alpha}{M_H^2}-\frac{\cos(\alpha-\beta)\sin\alpha}{M_h^2}\right)(F_{13}-F_{31}^*).
\end{split} \end{align}
    In Eq. (\ref{eq 0909.0863}), $v$, $\beta$, $\alpha$, $M_H$, $M_h$, $M_A$, $F_{13}$, and $F_{31}$ are vaccum expectation value, Higgs mixing angles, hevier nutral Higgs mass, lighter nutral Higgs mass, CP odd Higgs mass, and $b-d$ coupling constants, respectively, which are defined in Ref. \cite{arXiv:0909.0863}.
    This contribution can compete with or even dominate the SM one. 
    Especially, in $M_A\to\infty$ limit, $C_1$ contains only the SM effect and $C_2$ contains only the 2HDM contribution.

    On the other hand, the supersymmetric SM (SUSY) models without R-parity \cite{ph/9707371} suggest the coefficients
\begin{align} \begin{split}
C_1&=-\frac{\lambda^{*}_{k33} \lambda_{k3q}'+\lambda_{k33} \lambda'{}^{*}_{kq3}}{4G_F m_{\tilde l_k}^2}\biggr|_{(k\not =3)}-\frac{2m_\tau}{m_B} \left\{\frac{\lambda'{}^{*}_{3k3} \lambda_{3kq}'}{8G_F m_{\tilde q_k}^2}+\bigl[C_A^q\bigr]_{\mathrm{SM}} \right\}\\
C_2&=\frac{\lambda_{k33} \lambda'{}^{*}_{kq3}-\lambda^{*}_{k33} \lambda_{k3q}'}{4G_F m_{\tilde l_k}^2}\biggr|_{(k\not =3)},
\end{split} \end{align}
    where $\lambda_{ijk}$ and $\lambda'_{ijk}$ are the coefficients of $\hat L_i \hat L_j \hat l_k^c$ and $\hat L_i \hat Q_j \hat d_k^c$ type couplings, respectively, 
    where $\hat L$, $\hat l^c$, $\hat Q$, and $ \hat d^c$ are the lepton doublet, lepton singlet, quark doublet, down-type quark singlet superfields, respectively; 
    $m_{\tilde l_i}$ and $m_{\tilde q_i}$ are slepton and squark masses, respectively.

    Moreover, leptquark models \cite{ph/9309310}, the topcolor-assisted technicolor model \cite{0901.3463}, and the Babu-Kolda model \cite{ph/9909476} also deform the SM energy distribution,
    while the energy distribution in the multiscale walking technicolor model \cite{ph/9903345} is the same as the SM one.
    Considering the ratio $Br(B\to \tau\tau)/Br(B\to\mu\mu)$, SUSY models without R-parity \cite{ph/9707371}, leptquark models \cite{ph/9309310}, and the topcolor-assisted technicolor model \cite{0901.3463} predictions differ from the SM one,
    while the Ref. \cite{arXiv:0909.0863}, the Babu-Kolda model \cite{ph/9909476}, and the multiscale walking technicolor model \cite{ph/9903345} predict the same value as the SM one. 
    These charasteristic features of models are available to distinguish them. 
    
     All of these models predict that the $Br(B\to \tau\tau)$ may become larger than the SM one.
 


\subsection{Example 2 - $\tau^+$ Decays into $\pi^+ \bar \nu_\tau$}
    We show here another example, in which $B^0$ decays into $\tau^+\tau^-$ and subsequently $\tau^+\to\pi^+ +\bar\nu_\tau$.
    In this case, we can set $G_1^a(y_a)= \delta(1-y_\pi)/y_\pi^2$, $G_2^a(y_a)=-\delta(1-y_\pi)/y_\pi$, $\lambda_a = 1$, and $\beta_a=\beta_b=\sqrt{1-4m_\tau^2/m_B^2}\equiv \beta$.
    Hence, the energy distribution is
\begin{align}\begin{split}  
\frac{1}{\Gamma}\frac{d\Gamma}{  d x_\pi   }  
= \frac{1 }{\beta } 
\biggl\{ 1
   - \frac{2Re[\widetilde{C}_1\widetilde{C}_2^*]}{ |\widetilde{C}_1|^2 +|\widetilde{C}_2|^2 \beta^2} (1-2x_\pi)  
 \biggr\} \theta[x_\pi-\frac{1-\beta}{2}]\theta[\frac{1+\beta}{2}-x_\pi].
\end{split} \end{align}

\begin{figure}[ht]
  \begin{center}
    \includegraphics[keepaspectratio=true,height=50mm]{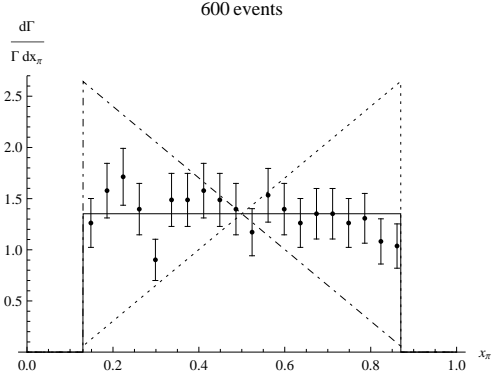}
  \end{center}
  \caption{The $\pi^+$ energy distribution of $B^0\to\tau^+ \tau^-$ and subsequently $\tau^+\to\pi^++\bar\nu_\tau$ or $\tau^-\to\pi^-+\nu_\tau$ decay, and their CP conjugate.
    The horizontal axis is the normalized $\pi^+$ energy $x_\pi$.
    The vertical axis is the time integrated differential decay width $d\Gamma/dx_\pi$ over the time integrated partial width $\Gamma$.
    We set $|{C}_1|=|{C}_2|=1$.
    The number of events to perform MC for $(Re[C_1 C_2^*]+Re[\bar C_1 \bar C_2^*])/|2C_1C_2|=0$ is 600.
 The solid line, dashed line, and dot-dashed line represent $(Re[C_1 C_2^*]+Re[\bar C_1 \bar C_2^*])/|2C_1C_2|=\{0,1,-1\}$ case, respectively.
  $(Re[C_1 C_2^*]+Re[\bar C_1 \bar C_2^*])/|2C_1C_2|=  -0.11\pm0.07$.        
   }
\label{1e-pi-c11-c21}
\end{figure}

\begin{figure}[ht]
  \begin{center}
    \includegraphics[keepaspectratio=true,height=50mm]{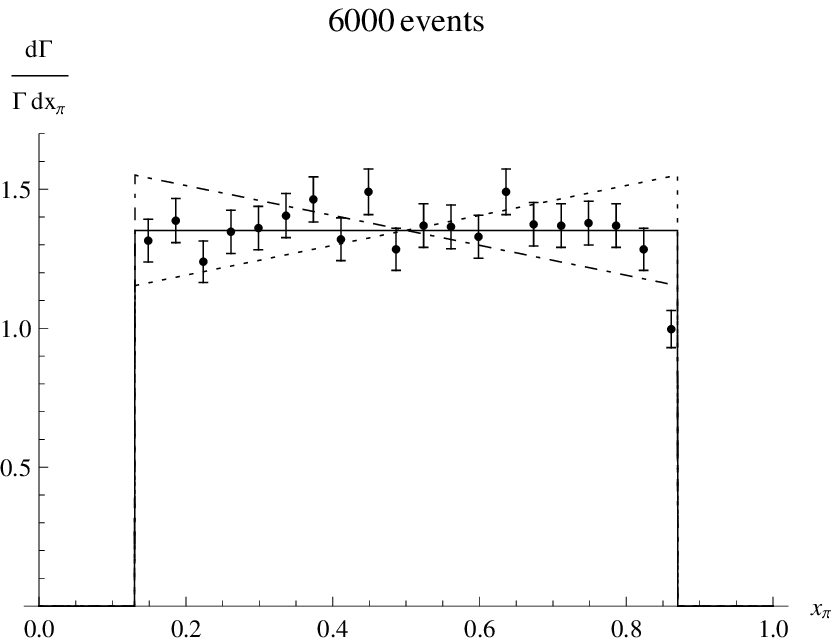}
  \end{center}
  \caption{The absolute values of coefficients are $|{C}_1|=1$ and $|{C}_2|=0.1$.
            The number of event is 6000. 
            Others are the same as Fig. \ref{1e-pi-c11-c21}.
  $(Re[C_1 C_2^*]+Re[\bar C_1 \bar C_2^*])/|2C_1C_2|= -0.21\pm0.15$. 
  }
\label{1e-pi-c11-c20.1}
\end{figure}

\begin{figure}[ht]
  \begin{center}
    \includegraphics[keepaspectratio=true,height=50mm]{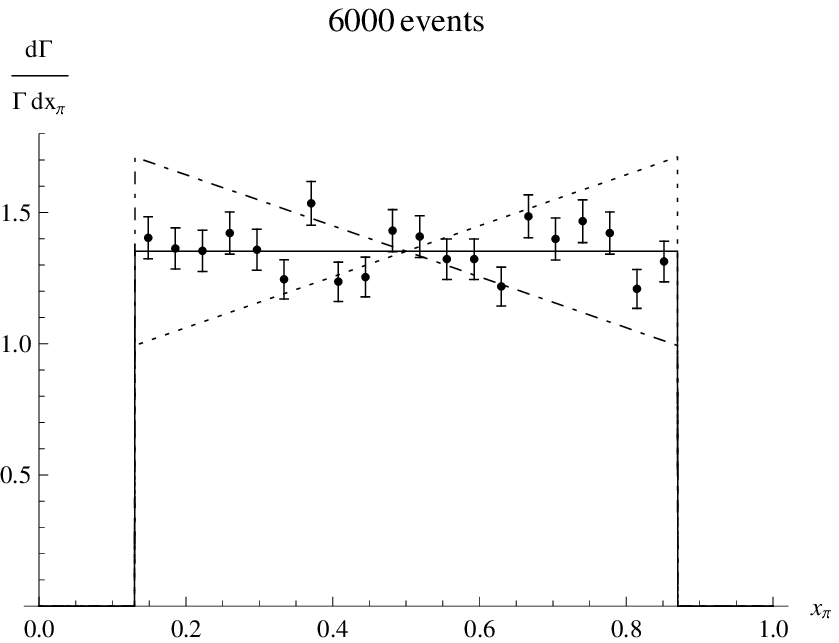}
  \end{center}
  \caption{The absolute values of coefficients are $|{C}_1|=0.1$ and $|{C}_2|=1$.
            The number of event is 6000. 
            Others are the same as Fig. \ref{1e-pi-c11-c21}.
  $(Re[C_1 C_2^*]+Re[\bar C_1 \bar C_2^*])/|2C_1C_2|=   -0.036\pm0.085  $. 
}
\label{1e-pi-c10.1-c21}
\end{figure}

    In Figs. \ref{1e-pi-c11-c21}-\ref{1e-pi-c10.1-c21}, we depict the time integrated distributions and the MC results for $(Re[C_1 C_2^*]+Re[\bar C_1 \bar C_2^*])/|2C_1C_2|=0$ as the previous example.
    Fig. \ref{1e-pi-c11-c21} represents the $|{C}_1|=|{C}_2|=1$ case. 
    Similarly, Fig. \ref{1e-pi-c11-c20.1} and Fig. \ref{1e-pi-c10.1-c21} represent the $\{|{C}_1|,|{C}_2|\}=\{1,0.1\}$ and $\{|{C}_1|,|{C}_2|\}=\{0.1,1\}$ cases, respectively.
    This mode is more suitable to understand the $B^0 \to \tau^+\tau^-$ current structure than preceding one, since two-body decay does not dilute the polarization unlike the previous case, even though the $\tau^+\to\pi^++\bar\nu_\tau$ branching ratio is about 0.11, which is smaller than the previous case.

    The results of MC are as follows:
\begin{align} \begin{split}
&\frac{Re[C_1 C_2^*]+Re[\bar C_1 \bar C_2^*]}{2|C_1C_2|}
\\&=
\left\{
  \begin{array}{lll}
  -0.11 \pm0.07       & \mathrm{for\ } |C_1|=|C_2|=1,\       &\mathrm{600\ events}   \\
  -0.21 \pm0.15       & \mathrm{for\ } |C_1|=1,\ |C_2|=0.1,\ &\mathrm{6000\ events}   \\
  -0.036\pm0.085      & \mathrm{for\ } |C_1|=0.1,\ |C_2|=1,\ &\mathrm{6000\ events}   \\
  \end{array}
\right. 
\end{split} \end{align}
  



\section{Opening Angle Distribution }\label{Opening Angle}

    Here, we consider the opening angle $\Theta$ between particles $a$ and $b$ in $B$ rest frame.
    The prescription is similar as Section \ref{Energy Distribution}, however, this time we multiply the different delta function
\begin{align} \begin{split}
 \delta\Bigl[\cos\Theta
 -\frac{\sin\theta_a \sin\theta_b \cos\phi
-\gamma_a\gamma_b (\beta_a+\cos\theta_a)
             (\beta_b-\cos\theta_b  )}
 {(1+\beta_a\cos\theta_a)(1-\beta_b\cos\theta_b)\gamma_a\gamma_b}
\Bigr].
\end{split} \end{align}
    Then, the result is
\begin{align}\begin{split}  \label{O-A}
&\frac{1}{\Gamma}\frac{d\Gamma}{ d\cos\Theta} 
\\ &=   
\frac{1}{4\pi }
\in -1 1  d\cos\theta_a \in {B_{min}(\Theta,\theta_a)} {B_{Max}(\Theta,\theta_a)}   d\cos\theta_b 
\\&\hspace{1em}\times
\frac{(1+\beta_a\cos\theta_a)(1-\beta_b\cos\theta_b)\gamma_a\gamma_b
}{\sqrt{\sin^2\theta_a \sin^2\theta_b-\gamma_a^2\gamma_b^2\{(1+\beta_a\cos\theta_a)(1-\beta_b\cos\theta_b)\cos\Theta
+ (\beta_a+\cos\theta_a)(\beta_b-\cos\theta_b  )
\}^2}}
\\&\hspace{1em}\times\Bigl\{1  
+\frac{D_4}{D_1}(\frac{m_b}{m_a}  \cos\theta_a \langle G_2^a(y_a)\rangle  +\frac{m_a}{m_b} \cos\theta_b \langle G_2^b(y_b)\rangle )
\\&\hspace{3em}-[\frac{D_2}{D_1} \gamma_a\gamma_b \{(1+\beta_a\cos\theta_a)(1-\beta_b\cos\theta_b)\cos\Theta
+ (\beta_a+\cos\theta_a)(\beta_b-\cos\theta_b  )
\}
\\&\hspace{4.5em}- \cos\theta_a \cos\theta_b
]\langle G_2^a(y_a)\rangle \langle G_2^b(y_b)\rangle \Bigr\},            
\end{split} \end{align}
    where we set
\begin{align} \begin{split}
  \int dy_a \frac{y_a^2}{\lambda_a} G^a_2(y_a)&\equiv \langle  G^a_2(y_a) \rangle,
\\\int dy_b \frac{y_b^2}{\lambda_b} G^b_2(y_b)&\equiv \langle  G^b_2(y_b) \rangle,
\end{split} \end{align}
\begin{align} \begin{split}
&B_{Max,min}(\Theta,\theta_a) 
\\=&
\frac{\gamma_a^2\gamma_b^2\{(1+\beta_a \cos \theta_a)\beta_b \cos\Theta+(\beta_a+\cos\theta_a)\}
              \{(1+\beta_a \cos \theta_a)      \cos\Theta+(\beta_a+\cos\theta_a)\beta_b\}
      }
     {\gamma_a^2\gamma_b^2\{(1+\beta_a \cos \theta_a)\beta_b \cos\Theta+(\beta_a+\cos\theta_a)\}^2
     +\sin^2\theta_a}
\\&\pm
\frac{\sin\theta_a
      \sqrt{\sin^2\theta_a+\gamma_a^2\{(\beta_a+\cos\theta_a)^2-(1+\beta_a\cos\theta_a)^2\cos^2\Theta\}
      }}
     {\gamma_a^2\gamma_b^2\{(1+\beta_a \cos \theta_a)\beta_b \cos\Theta+(\beta_a+\cos\theta_a)\}^2
     +\sin^2\theta_a}.
\end{split} \end{align}
    This expression suggests that the opening angle distribution determines $|C_1|$ and $|C_2|$, separately, via the coefficients $D_2$. 

    If $f_a=f_b$, and the decay modes of $\bar f_a$ and $f_b$ are the same, for example, $B^0\to \tau^+\tau^-\to \pi^+\pi^- \bar \nu_\tau\nu_\tau$ mode, the second term in Eq. (\ref{O-A}) which has the coefficient $D_4/D_1$ will vanish because this term is antisymmetric about the $\cos\theta_a + \cos\theta_b=0$ line, on the other hand, the domain of integration is symmetric.

\subsection{Example 3 - $\tau^\pm$ Decay into $\mu^\pm$}

\begin{figure}[ht]
  \begin{center}
    \includegraphics[keepaspectratio=true,height=50mm]{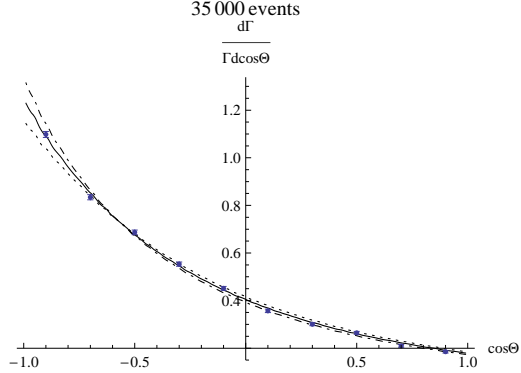}
  \end{center}
  \caption{The opening angle distribution between $\mu^+$ and $\mu^-$ in $B^0\to\tau^+ \tau^-\to\mu^++\mu^-+\nu_\tau \bar \nu_\tau \nu_\mu \bar \nu_\mu$ decay.
    The horizontal axis is $\cos\Theta$, and the vertical axis is time integrated $d\Gamma/ d\cos\Theta$ over time integrated $\Gamma $.  
   The solid line, dashed line, and dot-dashed line represent $(|C_1|^2-|C_2|^2\beta^2)/(|C_1|^2+|C_2|^2\beta^2)=\{0,1,-1\}$ cases, respectively.
   The MC result for $(|C_1|^2-|C_2|^2\beta^2)/(|C_1|^2+|C_2|^2\beta^2)=0$ is $(|C_1|^2-|C_2|^2\beta^2)/(|C_1|^2+|C_2|^2\beta^2)=-0.15\pm0.18$ in a sample of 35000 events.
}
\label{oa-mu}
\end{figure}

    We here show a simple example that $B^0$ decays into $\tau^+\tau^-$, and subsequently, they decay into $\mu^++\mu^-+\nu_\mu+\bar\nu_\mu+\nu_\tau+\bar\nu_\tau$.   
    In this case, we set  $\langle G_2^a(y_a) \rangle=\langle G_2^b(y_b) \rangle=1/3$.
    The numerical result is depicted in Fig. \ref{oa-mu}.
    We perform the MC for $(|C_1|^2-|C_2|^2\beta^2)/(|C_1|^2+|C_2|^2\beta^2)=0$ in a sample of 35000 events, which corresponds to the 100 times of 50 ab${}^{-1}$.
    The result is $(|C_1|^2-|C_2|^2\beta^2)/(|C_1|^2+|C_2|^2\beta^2)=-0.15\pm0.18$.

    In this figure, the increase near $\cos \Theta = -1$ is caused by the back-to-back Lorentz boost of $\bar f_a$ and $f_b$ along the $z$ axis.


    If new physics affects this mode substantially, we may detect the distribution.  
    In that case, $B\to\tau^+\tau^-$ opening angle distribution is usefull to distinguish new physics models. 
    In the SM, $|C_1|\gg |C_2|$ \cite{PRD55.2768}.
    Then, the shape of distribution is the same as $(|C_1|^2-|C_2|^2\beta^2)/(|C_1|^2+|C_2|^2\beta^2)=+1$ case.
    However, many new physics models, for example, Refs. \cite{arXiv:0909.0863}, \cite{ph/9707371}, \cite{ph/9309310}, \cite{0901.3463}, and \cite{ph/9909476}, deform the shape of distribution near $\cos \Theta = -1$.


\subsection{Example 4 - $\tau^\pm$ Decay into $\pi^\pm$}

\begin{figure}[ht]
  \begin{center}
    \includegraphics[keepaspectratio=true,height=50mm]{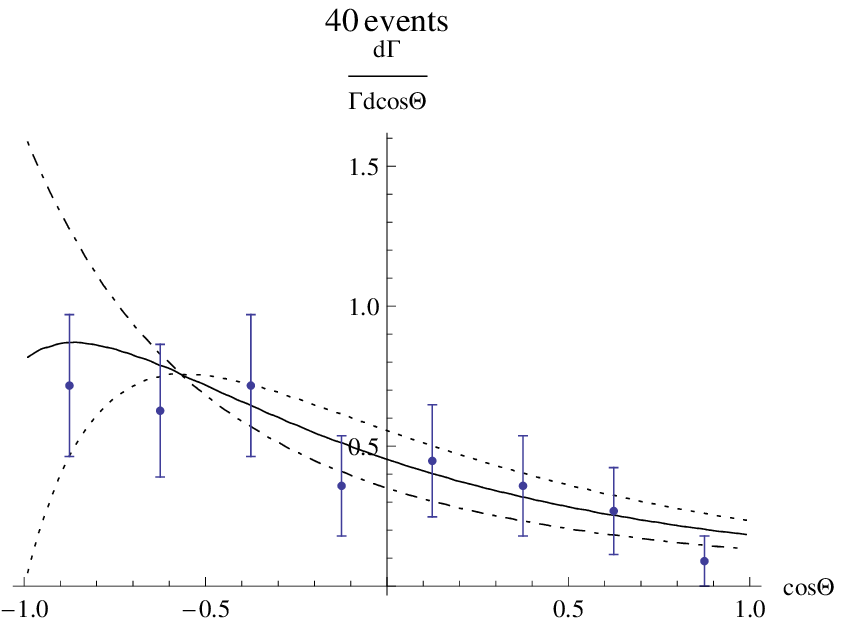}
  \end{center}
  \caption{The opening angle distribution between $\pi^+$ and $\pi^-$ in $B^0\to\tau^+ \tau^-\to\pi^++\pi^-+ \nu_\tau \bar \nu_\tau$ decay.
    The horizontal axis is $\cos\Theta$, and the vertical axis is $d\Gamma/(\Gamma d\cos\Theta)$.  
    The solid line, dashed line, and dot-dashed line represent $(|C_1|^2-|C_2|^2\beta^2)/(|C_1|^2+|C_2|^2\beta^2)=\{0,1,-1\}$ cases, respectively. 
    The MC result for $(|C_1|^2-|C_2|^2\beta^2)/(|C_1|^2+|C_2|^2\beta^2)=0$ is $(|C_1|^2-|C_2|^2\beta^2)/(|C_1|^2+|C_2|^2\beta^2)=0.03\pm0.49$ in a sample of 40 events.
}
\label{oa-pi}
\end{figure}

    We show another example that $B^0$ decays into $\tau^+\tau^-$, and subsequently, they decay into $\pi^++\pi^-+\nu_\tau+\bar\nu_\tau$.   
    In this case, we set $\langle G_2^a(y_a) \rangle=\langle G_2^b(y_b) \rangle=1$.
    The numerical result is depicted in Fig. \ref{oa-pi}.
    We perform the MC for $(|C_1|^2-|C_2|^2\beta^2)/(|C_1|^2+|C_2|^2\beta^2)=0$ in a sample of 40 events, which will given in the Super B-factory.
    The MC result is $(|C_1|^2-|C_2|^2\beta^2)/(|C_1|^2+|C_2|^2\beta^2)=0.03\pm0.49$.

    In this figure, the dashed line decreases near $\cos\Theta=-1$, on the other hand, the dot-dashed line increases there.
    The reason is as follows:
    In this case, the second term in Eq. (\ref{O-A}) is vanished, and we can set $\beta_a=\beta_b\equiv \beta$.
    Therefore, for $\cos\Theta=-1$, Eq. (\ref{O-A}) is proportional to 
\begin{align} \begin{split}
\Bigl\{(1  +\frac{D_2}{D_1})  
(1+\cos\theta_a\cos\theta_b) \Bigr\}.            
\end{split} \end{align}
    The factor $1  +D_2/D_1$ becomes zero when ${C}_2$ is zero, and it becomes maximum when ${C}_1$ is zero.

\section{Azimuthal Angle Asymmetry}

    Generally, the trajectories of $a$ and $b$ draw the skew lines since $\bar f_a$ and $f_b$ have the finite lifetimes.
    If the vertex detector of B-factory could detect the decay points of $\bar f_a$ and $f_b$,
     we were able to determine $\phi$ dependence of $d\Gamma$, and then $Im[C_1C_2^*]$.
    However, some of $\bar f_a$ and $f_b$ decay into one-prong modes, the polarization effect is diluted in the many body decays, and/or the vertex detector does not have sufficient resolution to detect the decay points accurately.
    Thus, we consider another method to determine $Im[C_1C_2^*]$.
    
    Since $\phi$ is the azimuthal angle between $a$ and $b$ as depicted in Fig. \ref{Frame-3}, the Lorentz boost along $z$ direction has no effect on this angle. 
    Thus, the delta function is unnecessary unlike the Sections \ref{Energy Distribution} and \ref{Opening Angle}.

    The trajectories of $a$ and $b$ in $B$ rest frame are written as
\begin{align} \begin{split}
   \mathbf{q}_a(t_a)&=\{\sin\theta_a t_a,0,(\beta_a\gamma_a+\gamma_a\cos\theta_a)t_a+d_z\}
\\ \mathbf{q}_b(t_b)&=\{\sin\theta_b\cos\phi t_b,\sin\theta_b\sin\phi t_b,(-\beta_b\gamma_b+\gamma_b\cos\theta_b)t_b\},
\end{split} \end{align}
    where $t_a$ and $t_b$ are the parameters.

\begin{figure}[htbp]
  \begin{center}
    \includegraphics[keepaspectratio=true,height=40mm]{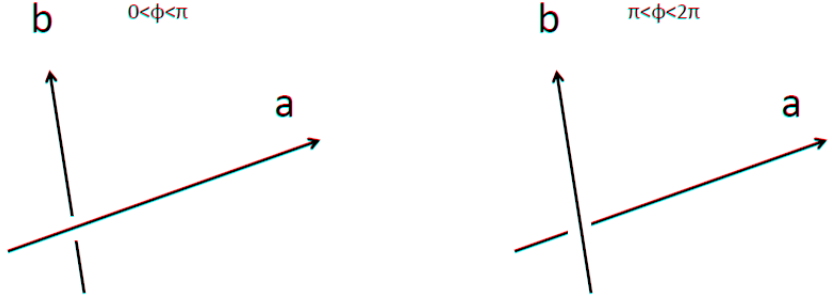}
  \end{center}
  \caption{Interpolating the trajectories, the skew lines take two types of alignments.
  The left one corresponds to $0<\phi<\pi$. 
  The right one corresponds to $\pi<\phi<2\pi$. 
  } 
  \label{fig:A-phi.eps}
\end{figure}

    The vector product of $\mathbf{q}_a(t_a)$ and $\mathbf{q}_b(t_b)$ for $d_z\to0$, $t_a>0$, and $t_b>0$  takes the form    
\begin{align} \begin{split}\label{V-P}
 &\mathbf{q}_a(t_a)\times \mathbf{q}_b(t_b)\bigr|_{d_z\to0,t_a>0,t_b>0}
 \\&=t_a t_b
 \left(
  \begin{array}{c}
 -\gamma_a(\beta_a+\cos\theta_a)\sin\theta_b\sin\phi \\
 \gamma_a(\beta_a+\cos\theta_a)\sin\theta_b\cos\phi-\sin\theta_a\gamma_b(-\beta_b+\cos\theta_b) \\
   \sin\theta_a\sin\theta_b\sin\phi \\
  \end{array}
\right).
\end{split} \end{align}
    Meanwhile, the difference between $\mathbf{q}_a(t_a')$ and $\mathbf{q}_b(t_b')$ is 
\begin{align} \begin{split}\label{a-b}
   \mathbf{q}_a(t_a')-\mathbf{q}_b(t_b')&
   =
 \left(
  \begin{array}{c}
   \sin\theta_a t_a'-\sin\theta_b\cos\phi t_b'\\
   -\sin\theta_b\sin\phi t_b'\\
   \gamma_a(\beta_a+\cos\theta_a)t_a'+d_z-\gamma_b(-\beta_b+\cos\theta_b)t_b'
  \end{array}
\right),
\end{split} \end{align}
    where $t_a'$ and $t_b'$ are the parameters of $\mathbf{q}_a(t_a')$ and $\mathbf{q}_b(t_b')$.
    $t_a'$ and $t_b'$ take arbitrary values.
    The scalar product between Eq. (\ref{V-P}) and Eq. (\ref{a-b}) is given by 
\begin{align} \begin{split}
 \mathbf{q}_a(t_a)\times \mathbf{q}_b(t_b)|_{d_z\to0,t_a>0,t_b>0}
 \cdot
 \bigl\{ \mathbf{q}_a(t_a')-\mathbf{q}_b(t_b')\bigr\}
 =t_a t_b
    d_z \sin\theta_a\sin\theta_b  \ \sin\phi.
\end{split} \end{align}
    This quantity becomes plus as $0<\phi<\pi$ and minus as $\pi<\phi<2\pi$.
    The sign of $\sin\phi$ is determined event-by-event (See Fig. \ref{fig:A-phi.eps}). 
    Then, the azimuthal angle asymmetry
\begin{align}\begin{split}  
\in {0} {\pi} d\phi \frac{1}{\Gamma} \frac{d\Gamma}{  d\phi} -\in {\pi} {2\pi} d\phi \frac{1}{\Gamma}\frac{d\Gamma}{  d\phi}  
=  - \frac{\pi}{8  }
\frac{D_5}{D_1} \langle G_2^a(y_a)\rangle \langle G_2^b(y_b)\rangle       
\end{split} \end{align}
    gives us the coefficient $D_5$, which is proportional to $Im[C_1C_2^*]$.

    We perform the MC for $B^0\to \tau^++\tau^-$ and then $\tau^+\to \pi^+ \bar\nu_\tau$ and $\tau^-\to \pi^- \nu_\tau$.
    We set $(Im[C_1 C_2^*]+Im[\bar C_1 \bar C_2^*])/|2C_1C_2|=0$ and summarize the results in table 1.
    
\begin{table}[htbp]
 \caption{The MC for $(Im[C_1 C_2^*]+Im[\bar C_1 \bar C_2^*])/|2C_1C_2|=0$ and  some conditions }
 \begin{center}
  \begin{tabular}{|c|c|c|}
    \hline
  $|C_1|$, $|C_2|$           & number of events   & $\frac{Im[C_1 C_2^*]+Im[\bar C_1 \bar C_2^*]}{2|C_1C_2|}$  and error \\
    \hline
    \hline
  $|C_1|=|C_2|=1$            & 40                 &  $-0.40\pm0.64$   \\
    \hline
  $|C_1|=|C_2|=1$            & 400                &  $-0.053\pm0.19$   \\
    \hline
  $|C_1|=1$, $|C_2|=0.1$     & 4000               &  $0.078\pm0.39$   \\
    \hline
  \end{tabular}
 \end{center}
\end{table}





\section{Only One of Two Fermions is Unstable}\label{only one}

    If $f_b$ is a stable particle, for example, $B_d^0\to \bar \Lambda_c^- p$, $B_u^+ \to \bar\Sigma_c(2455)^0 p$, $B^+\to\tau^+ \nu_\tau$, $B^0\to\tau^+e^-$, and $B^0\to\tau^+\mu^-$, the general formula (\ref{master-formula}) is modified to form
\begin{align}\begin{split} \label{master-formula-for one}
\frac{d\Gamma}{  dy_a d\Omega_a } 
 =    
Br_a\frac{y_a^2}{4\pi  \lambda_a}
 \frac{f_B^2G_F^2}{2\pi}|\mathbf{p}|
\Bigl\{D_1  G_1^a(y_a)  
+D_4 \frac{m_b}{m_a}  \cos\theta_a G_2^a(y_a) 
\Bigr\}.            
\end{split} \end{align}
    Then, by the similar calculations, the partial decay width is
\begin{align}\begin{split} 
\Gamma =   Br_a \frac{f_B^2G_F^2}{2\pi}|\mathbf{p}| D_1,             
\end{split} \end{align}
    the $a$ energy distribution is
\begin{align}\begin{split} \label{one-x}
\frac{1}{\Gamma}\frac{d\Gamma}{  dx_a } 
 =    \int dy_a  
\frac{1}{ \beta_a \lambda_a}
\Bigl\{ y_a G_1^a(y_a)  
+\frac{D_4}{D_1} \frac{m_b}{m_a\beta_a}(2x_a-y_a)  G_2^a(y_a) 
\Bigr\},            
\end{split} \end{align}
    where $\int dy_a$ means the same as before,
    and the distribution of the opening angle $\Theta'$ between $f_b$ and $a$ is
\begin{align}\begin{split} \label{one-Theta}
\frac{1}{\Gamma}\frac{d\Gamma}{ d\cos\Theta' }
 =   
\frac{1}{2  }
 \frac{1-\beta_a^2}{(1+\beta_a \cos\Theta')^2} 
\Bigl\{1-\frac{D_4}{D_1} \frac{m_b}{m_a}  \frac{\cos\Theta'+\beta_a}{1+\beta_a \cos\Theta'} \langle G_2^a(y_a)\rangle 
\Bigr\},            
\end{split} \end{align}
    where
\begin{align} \begin{split}
\cos\Theta'
=-\frac{\beta_a + \cos\theta_a}{1+\beta_a \cos\theta_a}.
\end{split} \end{align}
    Both of these two distributions give $D_4$.
    These are used for a cross-check. 
    However, we cannot pull out $D_2$ and $D_5$.
    The energy distribution is useful even if $f_b$ is a missing fermion except for neutrinos.
    If $f_b$ is a neutrino, $m_b\to 0$ and the second terms in both of Eqs. (\ref{one-x}) and (\ref{one-Theta}) are vanish, and then we cannot determine $D_4$.

\subsection{Example 5 - $B^0\to\tau^+\mu^-$,$\tau^+\to\pi^+ \bar\nu_\tau$}

    We consider the lepton flavor violating $B^0\to\tau^+\mu^-$ decay, and subsequently $\tau^+$ decays into $\pi^+ \bar\nu_\tau$.  
    In this case, we can set $G_1^a(y_a)= \delta(1-y)/y^2$, $G_2^a(y_a)=-\delta(1-y)/y$, $\lambda_a = 1$, and $\langle G_2^a(y_a) \rangle=1$.

    The $a$ energy distribution is
\begin{align}\begin{split} 
\frac{1}{\Gamma}\frac{d\Gamma}{  dx_a } 
=\frac{1}{ \beta_\tau }
\Bigl\{ 1  
-\frac{D_4}{D_1} \frac{m_\mu}{m_\tau\beta_\tau}(2x_\tau-1) \Bigr\}\theta[x_\tau-\frac{1-\beta_\tau}{2}]\theta[\frac{1+\beta_\tau}{2}-x_\tau].            
\end{split} \end{align}

\begin{figure}[ht]
  \begin{center}
    \includegraphics[keepaspectratio=true,height=50mm]{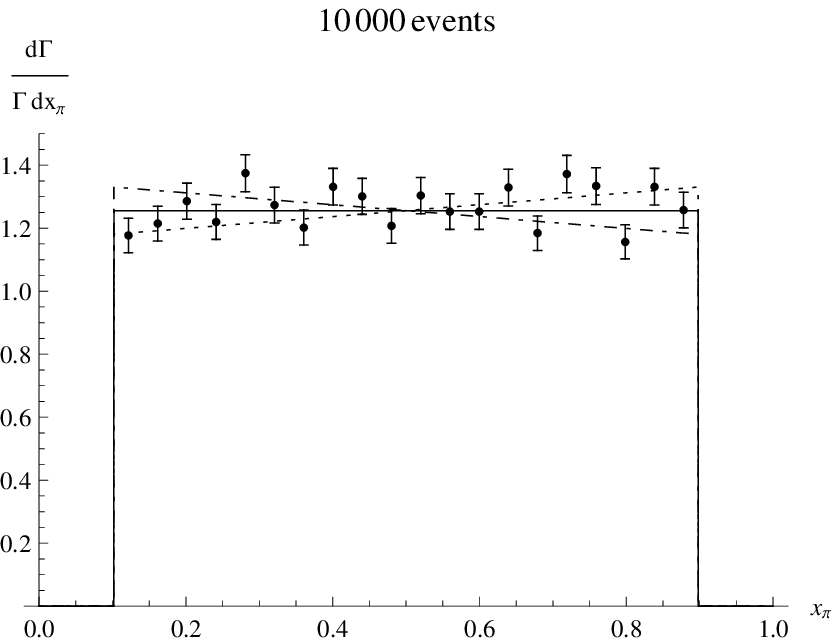}
  \end{center}
  \caption{The energy distribution of $\pi^+$ in $B^0\to\tau^+ \mu^-$, $\tau^+\to\pi^+ \bar\nu_\tau$ for $|{C}_1|=| {C}_2|=1$.
    The solid line, the dashed line, and the dot-dashed line represent $(Re[{C}_1 {C}_2^*]+Re[\bar{C}_1 \bar{C}_2^*])/|2{C}_1{C}_2|=\{0,1,-1\}$, respectively.
    The MC is performed for $(Re[{C}_1 {C}_2^*]+Re[\bar{C}_1 \bar{C}_2^*])/|2{C}_1{C}_2|=0$ in a sample of 10000 events to result in $(Re[{C}_1 {C}_2^*]+Re[\bar{C}_1 \bar{C}_2^*])/|2{C}_1{C}_2|=0.27\pm0.29$.}
  \label{taumu-x}
\end{figure}

\begin{figure}[ht]
  \begin{center}
    \includegraphics[keepaspectratio=true,height=50mm]{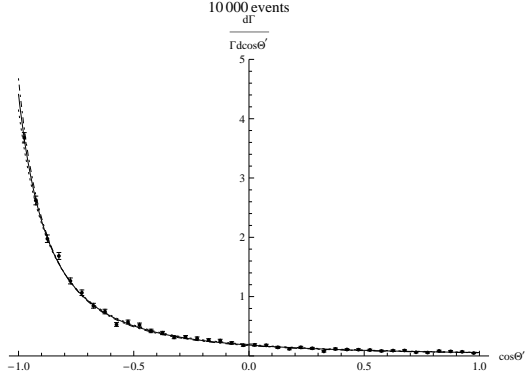}
  \end{center}
  \caption{The distribution of opening angle between $\mu^-$ and $\pi^+$ in the same process as Fig. \ref{taumu-x}. 
    The solid line, the dashed line, and the dot-dashed line represent $(Re[{C}_1 {C}_2^*]+Re[\bar{C}_1 \bar{C}_2^*])/|2{C}_1{C}_2|=\{0,1,-1\}$, respectively.
    The MC is performed for $(Re[{C}_1 {C}_2^*]+Re[\bar{C}_1 \bar{C}_2^*])/|2{C}_1{C}_2|=0$ in a sample of 10000 events to result in $(Re[{C}_1 {C}_2^*]+Re[\bar{C}_1 \bar{C}_2^*])/|2{C}_1{C}_2|=-0.03\pm0.30$.  }
  \label{taumu-Thetap}
\end{figure}

    On the other hand, the opening angle distribution takes the form 
\begin{align}\begin{split} 
\frac{1}{\Gamma}\frac{d\Gamma}{ d\cos\Theta' }
 =   
\frac{1}{2  }
 \frac{1-\beta_\tau^2}{(1+\beta_\tau \cos\Theta')^2} 
 \Bigl\{1   
-\frac{D_4}{D_1} \frac{m_\mu}{m_\tau}  \frac{\cos\Theta'+\beta_\tau}{1+\beta_\tau \cos\Theta'}
\Bigr\}.            
\end{split} \end{align}
    These distributions and the MC for $(Re[{C}_1 {C}_2^*]+Re[\bar{C}_1 \bar{C}_2^*])/|2{C}_1{C}_2|=0$ in a sample of 10000 events are depicted in Figs. \ref{taumu-x} and \ref{taumu-Thetap}.  
    The results are $(Re[{C}_1 {C}_2^*]+Re[\bar{C}_1 \bar{C}_2^*])/|2{C}_1{C}_2|=0.27\pm0.29$ and $-0.03\pm0.30$, respectively.
    They have almost the same errors.

    The SUSY models without R-parity \cite{hep-ph/9806359} suggest the coefficients
\begin{align} \begin{split}
C_1&=-\frac{1}{4G_F}\left\{\frac{m_B}{m_b}(C_{2S}^++C_{1S}^+)-\frac{m_\tau}{m_B}C_{V}^+\right\}\\
C_2&=-\frac{1}{4G_F}\left\{\frac{m_B}{m_b}(C_{2S}^+-C_{1S}^+)-\frac{m_\tau}{m_B}C_{V}^+\right\},
\end{split} \end{align}
    where $C_{1S}^+$, $C_{2S}^+$, and $C_{V}^+$ are defined in Ref. \cite{hep-ph/9806359} as
\begin{align} \begin{split}
C_{1S}^+=\sum_{i\not =3} \frac{\lambda'{}^{*}_{i3q} \lambda_{i32}}{m_{\tilde \nu_i}^2}, \hspace{2em}
C_{2S}^+=\sum_{i\not =2} \frac{\lambda'_{iq3} \lambda^{*}_{i23}}{m_{\tilde \nu_i}^2}, \hspace{2em}
C_{V}^+=\sum_{i} \frac{\lambda'{}^{*}_{2iq} \lambda_{3i3}}{2m_{\tilde q_i}^2},
\end{split} \end{align}    
    where $m_{\tilde \nu_i}$ is squark mass.
    In this model, the branching ratio can become about $10^{-5}$, which is the same order as the experimental upper bound, and it has a relation, $Br(B\to \mu\mu)\siml Br(B\to \tau\mu)\siml Br(B\to \tau\tau)$.
    Moreover, since this mode has only one neutrino, the efficiency is much higher than that of $B\to \tau\tau$ mode.


\section{ Example 6 - $B_u^+ \to \bar \Xi_c^0 \Lambda_c^+ $ and $\bar B^0\to \Lambda_c^+ \bar p$ Decay}\label{baryon}

    Now we show two baryon modes for examples.
    The first one is $B_u^+ \to \bar \Xi_c^0 \Lambda_c^+ $ decay.
    According to Ref. \cite{ph/0512335}, in the SM, we have the relation for $B^+$
\begin{align} \begin{split}
\frac{C_1}{C_2}=-\frac{m_{\Xi_c}^2-(m_B-m_{\Lambda_c})^2}
                      {m_B^2-(m_{\Xi_c}-m_{\Lambda_c})^2}
= 0.10, 
\end{split} \end{align}
    where $m_{\Xi_c}$ and $m_{\Lambda_c}$ are $\Xi_c^0$ and $\Lambda_c^+$ masses, respectively.
    Then, we predict 
\begin{align} \begin{split}\label{result-5}
  \frac{D_2}{D_1}
 = 0.89
,\hspace{3em} \frac{D_4}{D_1}
= -0.45
,\hspace{3em} \frac{D_5}{D_1}= 0.
\end{split} \end{align}
    This prediction is available for the test of Ref. \cite{ph/0512335}. 
%
%


    The next one is $\bar B^0\to \Lambda_c^+ \bar p$ decay.
    According to the Ref. \cite{PRD46-466}, which uses the factorization, 
\begin{align} \begin{split}
\frac{|C_2|}{|C_1|}
=0.34.
\end{split} \end{align}
    Then, if $C_1 C_2^*$ has no relative phase (or $Re[C_1C_2^*]=|C_1C_2|$), 
\begin{align} \begin{split}\label{factorization}
\frac{D_4}{D_1}=-0.52.
\end{split} \end{align}

    On the other hand, according to Ref. \cite{jarfi and lazrak}, which uses the pole model,
\begin{align} \begin{split}
\frac{|C_2|}{|C_1|}
=0.77
\end{split} \end{align}
    in $\bar B^0\to \Lambda_c^+ \bar p$ decay.
    Therefore, if $C_1 C_2^*$ has no relative phase (or $Re[C_1C_2^*]=|C_1C_2|$), 
\begin{align} \begin{split}\label{pole model}
\frac{D_4}{D_1}=-0.90.
\end{split} \end{align}


    These two models suggest different branching ratio and $|C_2/C_1|$. 
    The different branching ratio may be corrected by the non-perturbative QCD effect.
    However, $|C_2/C_1|$ eminently represents the feature of each model.
    Hence, we can test which model works better.


    The QCD effect may pollute new physics effect.
    However, at least, one of these observables is measured to considerably differ from (\ref{result-5}), or (\ref{factorization}) and (\ref{pole model}), we should take into account new physics.
    We point out that more observables are desirable for new physics discovery.


    Theoretically, deriving the precise expressions of $G_1$ and $G_2$ is not easy.
    However, we are interested in $B\to \bar f_a f_b$ decays. 
    We assume that there is no new physics contribution in the $\Xi_c^0$ and $\Lambda_c^+ $ decays.
    So, it is not necessary for us to know the expressions of $G_1$ and $G_2$ theoretically if we can determine it from the experimental data.
    Actually, Refs. \cite{PDG1}, \cite{hep-ex/0509042}, \cite{hep-ex/9912003}, and \cite{hep-ex/0112025} suggest the $\Lambda_c$ polarization, also Ref. \cite{hep-ex/0011073} measures the $\Xi_c^0$ polarization.
    In this reference, $\Xi_c^0$ decays into $\Xi^-\pi^+$.
    The two body decay makes $G_1$ and $G_2$ trivially.
    Then, these decays are given by
\begin{align} \begin{split}
\frac{dBr( \bar \Xi_c^0 \to \pi^-+p^+)}{d^3k_{\pi^-}}  
&= \frac{2Br_{\bar \Xi_c^0}}{\pi m_{\pi^-}^3 }\left[ \frac{\delta(1-y_{\pi^-})}{y_{\pi^-}^2}-\alpha_{\Xi_c}  \mathbf{s}^{\bar \Xi_c^0} \cdot \hat {\mathbf{k}}_{\pi^-}   \frac{\delta(1-y_{\pi^-})}{y_{\pi^-}} \right]\\ 
\frac{dBr( \Lambda_c^+ \to \pi^++\Lambda)}{d^3k_{\pi^+}}  
&= \frac{2Br_{\Lambda_c^+}}{\pi m_{\pi^+}^3 }\left[ \frac{\delta(1-y_{\pi^+})}{y_{\pi^+}^2}+\alpha_{\Lambda_c}  \mathbf{s}^{\Lambda_c^+} \cdot \hat {\mathbf{k}}_{\pi^+}   \frac{\delta(1-y_{\pi^+})}{y_{\pi^+}} \right],
\end{split} \end{align}
    where $Br_{\bar \Xi_c^0}=Br( \bar \Xi_c^0 \to \pi^-+p^+)$ and $Br_{\Lambda_c^+}=Br( \Lambda_c^+ \to \pi^++\Lambda)$; 
    $m_{\pi^\pm}$ are the charged pion masses, $y_{\pi^\pm}$ are normalized charged pion energies defined by the same manner as in Eq. (\ref{yayb}); 
    $\alpha_{\Xi_c} $ and $\alpha_{\Lambda_c}$ are the decay parameters; 
    $ \mathbf{s}^{\bar \Xi_c^0}$ and $ \mathbf{s}^{\Lambda_c^+}$ are the polarization vectors of $\bar \Xi_c^0$ and $\Lambda_c^+$, respectively;
    $ \mathbf{k}_{\pi^\pm}$ are the $\pi^\pm$ momenta, respectively.

        










%
%
%
%

\begin{figure}[htbp]
  \begin{center}
    \includegraphics[keepaspectratio=true,height=50mm]{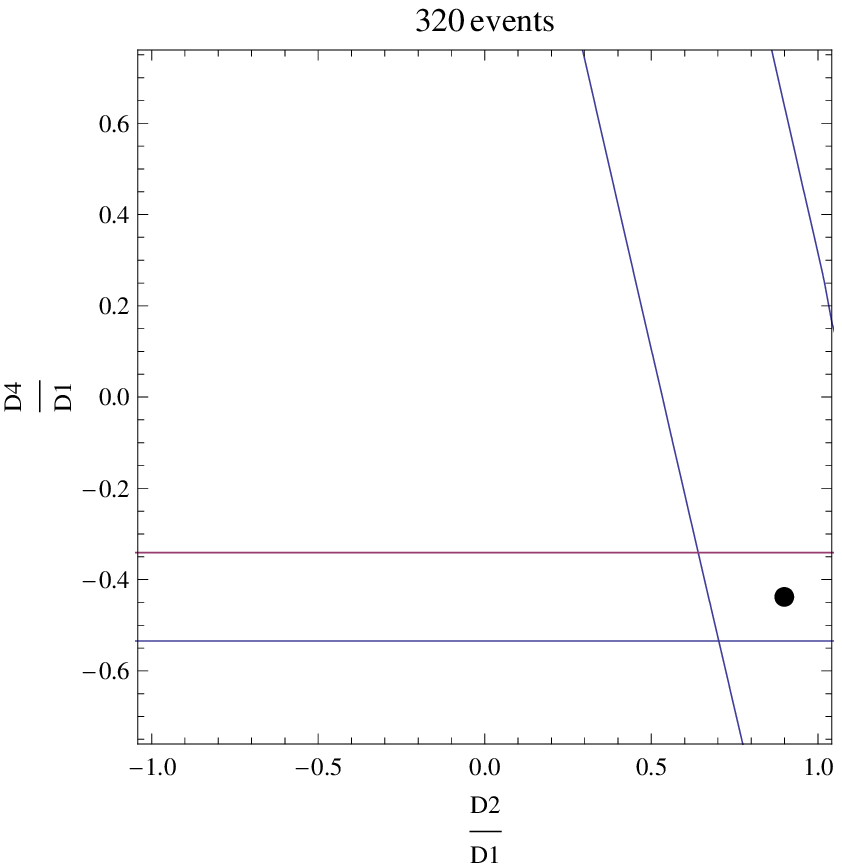}
    \includegraphics[keepaspectratio=true,height=50mm]{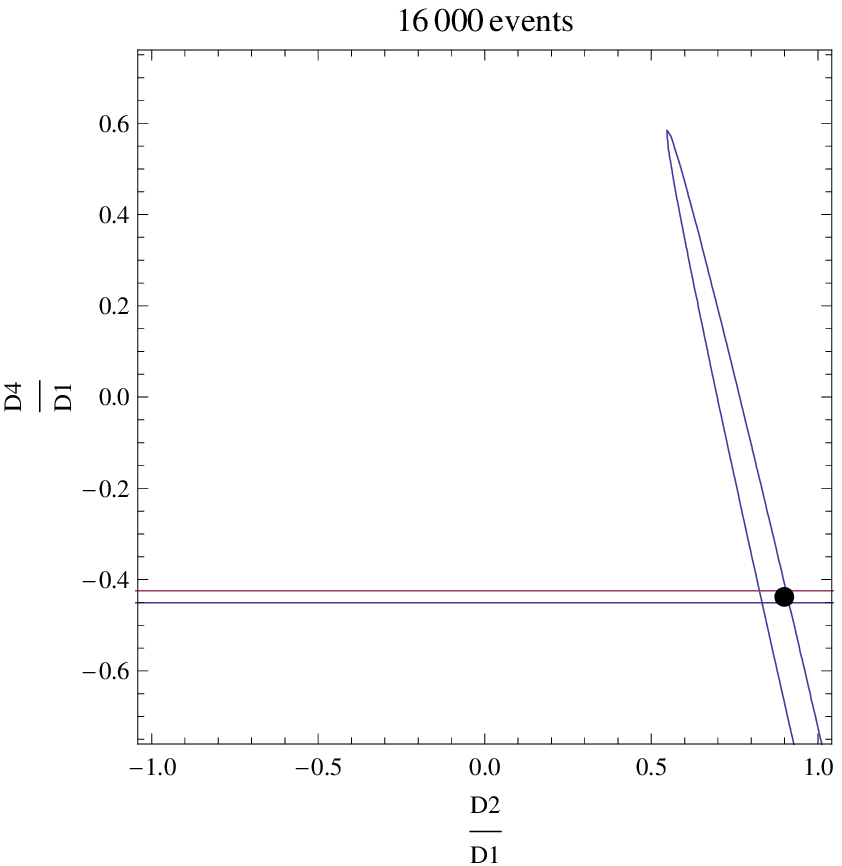}
  \end{center}
  \caption{The allowed region determined by the energy distribution and opening angle distribution of $B_u^+  \to \bar \Xi_c^0 + \Lambda_c^+ $, $\Lambda_c^+ \to \Lambda + \pi^+$, and $\bar \Xi_c^0 \to  p^+ + \pi^-$ decay chain.
    The left and right figures are the results of MC simulation with 320 and 16000 events, respectively.
    In each figure, the dot means the SM prediction. 
    The horizontal lines explain the allowed region which is determined by the energy distribution.
    The diagonal region is allowed by the opening angle distribution.
     $D_4/D_1=-0.44\pm0.10$ for 320 events. $D_4/D_1=-0.44\pm0.01$ for 16000 events. 
     }
  \label{fig:BtoSigma_c+Xi_c}
\end{figure}

\begin{figure}[htbp]
  \begin{center}
    \includegraphics[keepaspectratio=true,height=50mm]{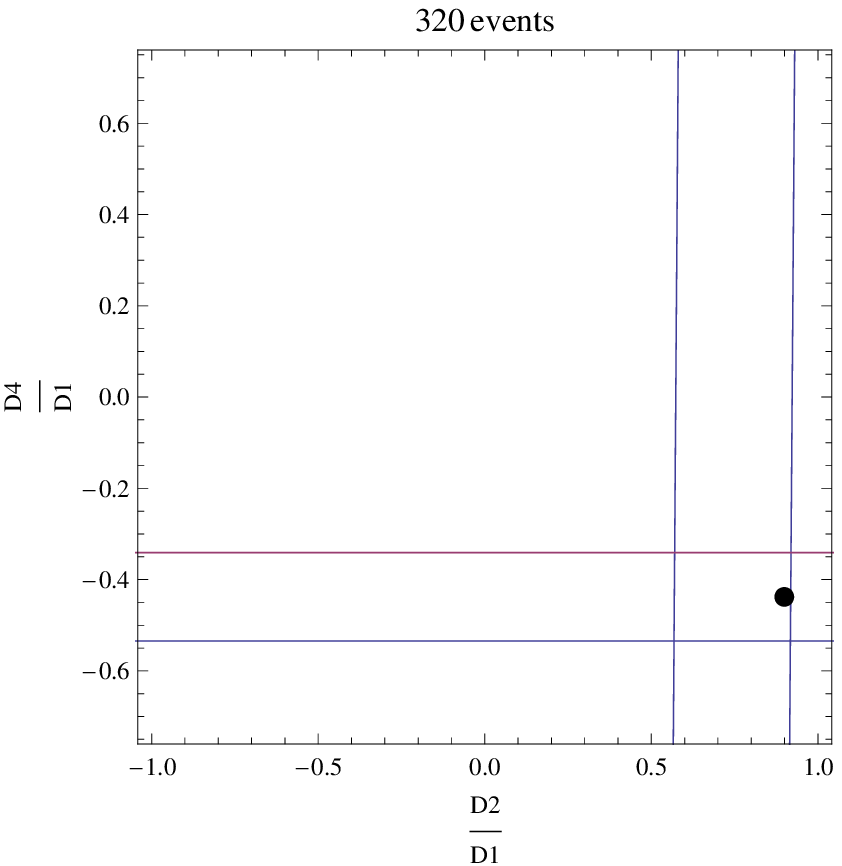}
    \includegraphics[keepaspectratio=true,height=50mm]{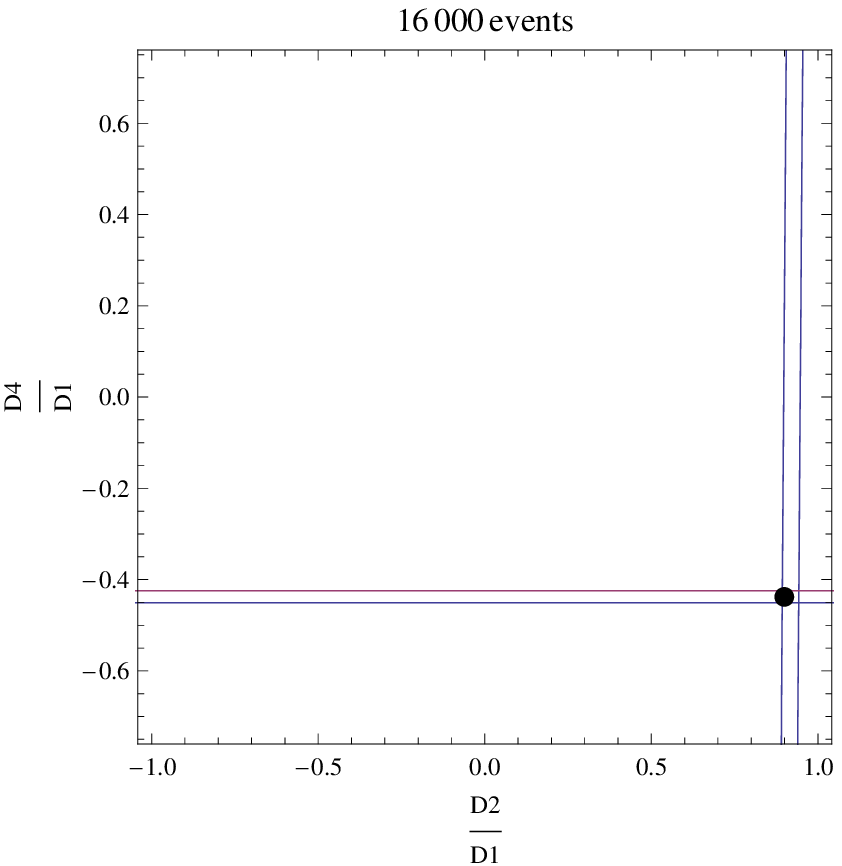}
  \end{center}
  \caption{Same as Fig. \ref{fig:BtoSigma_c+Xi_c} but, we here use $\alpha_{\Xi_c}=-0.6-0.4$.
    The diagonal region in Fig.  \ref{fig:BtoSigma_c+Xi_c} becomes narrower and more vertically.
     }
  \label{fig:BtoSigma_c+Xi_c-a=-1}
\end{figure}

\begin{figure}[htbp]
  \begin{center}
    \includegraphics[keepaspectratio=true,height=50mm]{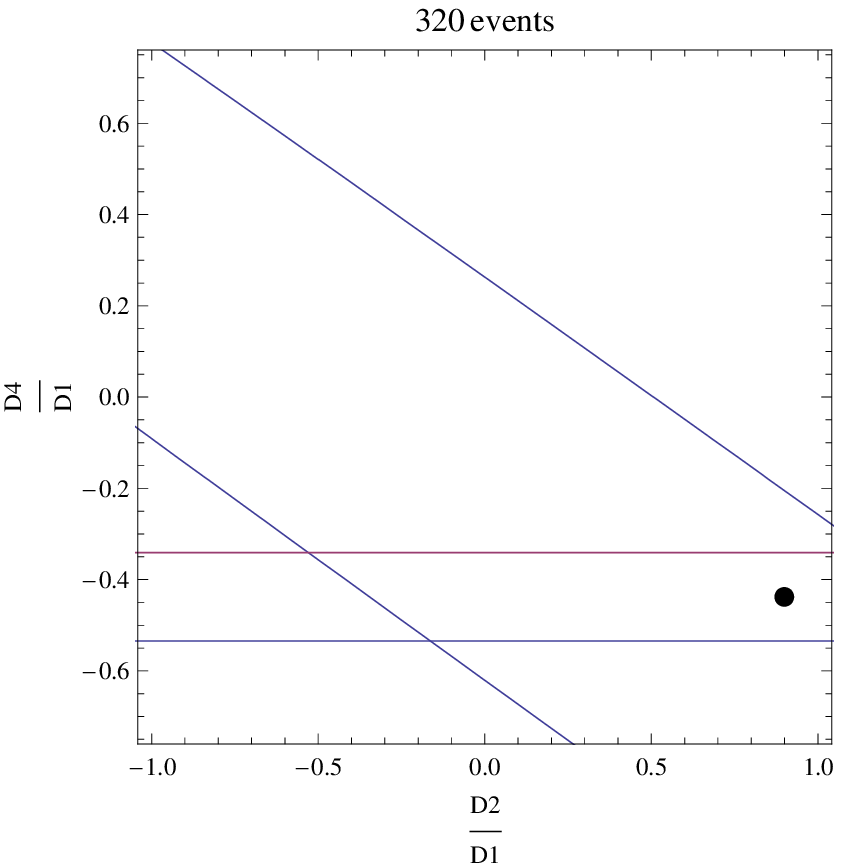}
    \includegraphics[keepaspectratio=true,height=50mm]{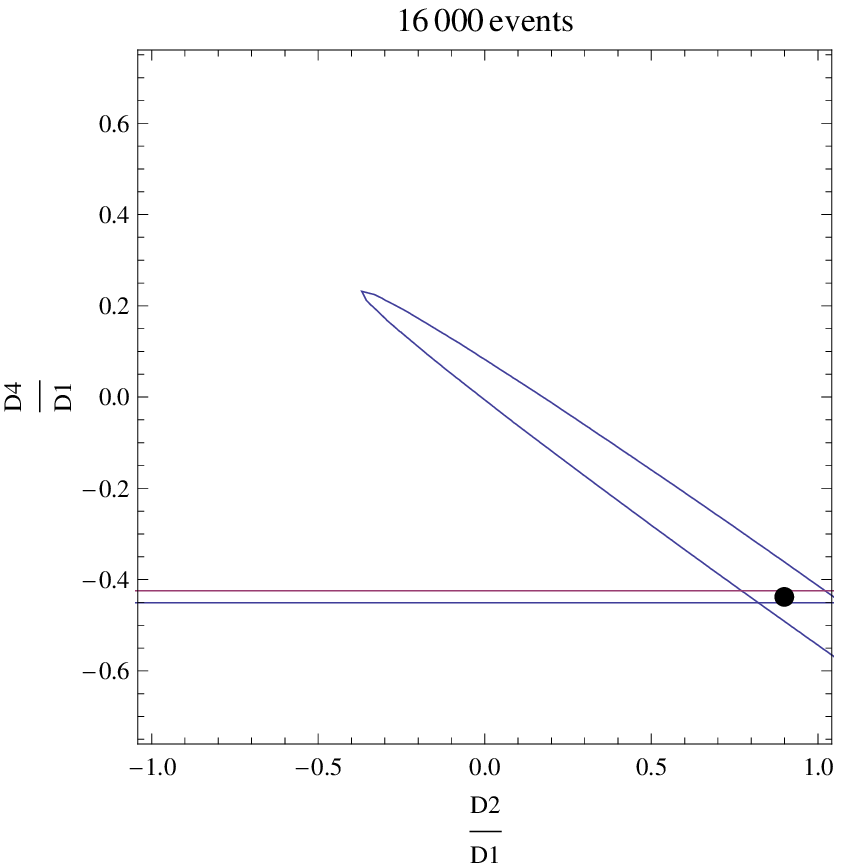}
  \end{center}
  \caption{Same as Fig. \ref{fig:BtoSigma_c+Xi_c} but, we here use $\alpha_{\Xi_c}=-0.6+0.4$.
    The diagonal region in Fig.  \ref{fig:BtoSigma_c+Xi_c} becomes wider and is inclined more horizontally.
     }
  \label{fig:BtoSigma_c+Xi_c-a=-02}
\end{figure}

    Fig, \ref{fig:BtoSigma_c+Xi_c} explains the $B_u^+  \to \bar \Xi_c^0 + \Lambda_c^+ $, $\Lambda_c^+ \to \Lambda + \pi^+$, and $\bar \Xi_c^0 \to  p^+ + \pi^-$ decay Monte Carlo simulation with 320 and 16000 events.
    The horizontal region is determined by the $\pi^+$ energy distribution 
    and the diagonal region is determined by the $\pi^+ \pi^-$ opening angle distribution.
    The dot represents Eq. (\ref{result-5}).
    The simulation is performed with this parameter set and the decay parameters, $\alpha_{\Xi_c}=-0.6$ and $ \alpha_{\Lambda_c}=-0.91$.
    We have to consider that $\alpha_{\Xi_c}$ has a large ambiguity $\alpha_{\Xi_c}=-0.6\pm0.4$.
    So, we explain $\alpha_{\Xi_c}$ dependence in Figs. \ref{fig:BtoSigma_c+Xi_c-a=-1} and \ref{fig:BtoSigma_c+Xi_c-a=-02}.



\begin{figure}[htbp]
  \begin{center}
    \includegraphics[keepaspectratio=true,height=45mm]{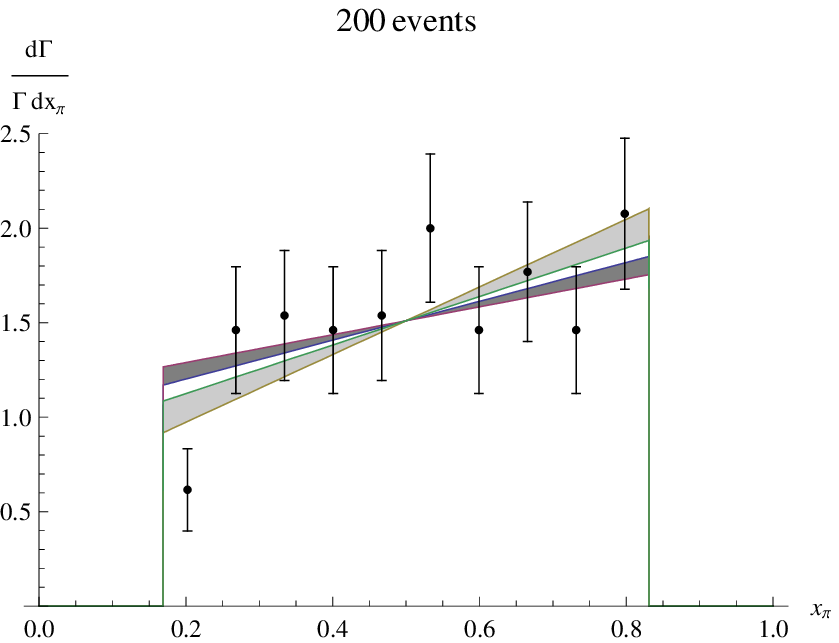}
    \includegraphics[keepaspectratio=true,height=45mm]{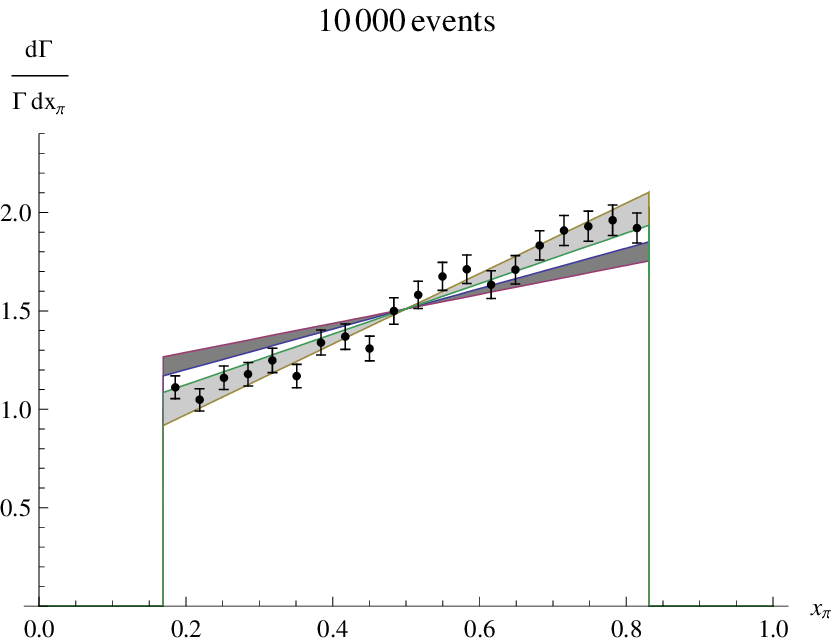}
  \end{center}
  \caption{
    The $\pi^+$ energy distribution of $\bar B^0\to \Lambda_c^+ \bar p$ and $\Lambda_c^+ \to \Lambda + \pi^+$ decay chain.
    The lighter gray explains $|C_2/C_1|=0.77$ case and the darker gray does $|C_2/C_1|=0.34$ case.
    These have breadth caused by the ambiguity in $\alpha_{\Lambda_c}=0.91\pm0.15$.   
    The dots with error bar are the result of MC simulation for  $|C_2/C_1|=0.77$ case with 200 (left figure) and 10000 (right figure) events.
    These suggest that $\alpha_{\Lambda_c}D_4/D_1 =0.72\pm0.27$ (left figure) and $\alpha_{\Lambda_c}D_4/D_1=0.83\pm0.04$ (right figure).
}
 \label{fig:Btop+Xi_c-Energy-10000.eps}
\end{figure}

    Fig. \ref{fig:Btop+Xi_c-Energy-10000.eps} represents the $\pi^+$ energy distribution of $\bar B^0\to \Lambda_c^+ \bar p$ and $\Lambda_c^+ \to \Lambda + \pi^+$ decay chain.
    The lighter gray explains $|C_2|/|C_1|=0.77$ case and the darker gray does $|C_2|/|C_1|=0.34$ case.
        These distributions have breadth caused by the ambiguity in $\alpha_{\Lambda_c}=-0.91\pm0.15$.   
        The results of Monte Carlo simulation with 200 and 10000 events are 
\begin{align} \begin{split}
\alpha_{\Lambda_c}\frac{D_4}{D_1}& =0.72\pm0.27, \\ 
\alpha_{\Lambda_c}\frac{D_4}{D_1}& =0.83\pm0.04,
\end{split} \end{align}
    respectively.

\section{Summary and Discussion}

    We studied the current structure of $B\to\bar f_a f_b$ decay modes using polarization effects.
    This can be applied to both of leptonic and baryonic decays, also, to the charged and neutral $B$ mesons.

    The $a$ energy distribution gives $Re[C_1 C_2^*]$.
    If we consider no or small relative phase between $C_1$ and $C_2^*$, we can estimate the ratio of $|C_1|$ and $|C_2|$.
    The energy distribution of $a$ and $b$ gives no more information.  
    
    The opening angle $\Theta$ distribution gives $|C_1|$ and $|C_2|$, separately.  
    With the energy distribution, this gives us the relative phase between $C_1$ and $C_2^*$ up to a binary ambiguity.  

    The azimuthal angle $\phi$ asymmetry gives $Im[C_1 C_2^*]$.
    We cannot detect the decay point in the one-prong events.
    Then, we cannot determine the $\phi$ distribution.
    However, we can determine that $\phi$ is larger or smaller than $\pi$. 
    This is enough to give $Im[C_1 C_2^*]$.

    If one of two fermions is stable particle, we cannot determine $D_2$ and $D_5$.
    However, $D_4$ is determined by each of the $\cos\Theta'$ and the $x_a$ distribution.

    We predicted $D_4/D_1$ and $D_2/D_1$ of the baryon modes $B_u^+ \to \bar \Xi_c^0 \Lambda_c^+ $ and $\bar B^0\to \Lambda_c^+ \bar p$.
    They are summarized in Figs. \ref{fig:BtoSigma_c+Xi_c}-\ref{fig:Btop+Xi_c-Energy-10000.eps}.






    In the Examples 1-4, we derived the $B^0_d\to\tau^+\tau^-$ sample number ignoring the efficiency.
    Here, we try to consider it.
    According to Ref. \cite{hep-ex/0511015}, they conclude $Br(B_d^0\to\tau^+\tau^-)<4.1\times10^{-3}$ using $(232\pm3)\times10^6$ data sample, which corresponds to $210$ $\mathrm{fb^{-1}}$.
    Hence, we need $7.9\times10^{12}$ ($7.2\times10^3$ $\mathrm{ab}^{-1}$) data sample to discover a $B^0_d\to\tau^+\tau^-$ event with the same efficiency as Ref. \cite{hep-ex/0511015}. 
    This efficiency can be improved, for example, by the semileptonic tagging method \cite{arXiv:0809.3834}.
    However, it is difficult to detect this mode in the SM case.
    If $B^0_d\to\tau^+\tau^-$ is detected, it must be induced by new physics.
    Then, our analysis is useful to determine its current structure.



    In the neutral $B$ decays, $\widetilde{C}_1$ and $\widetilde{C}_2$ are the functions of $t$ as defined in Eq. (\ref{td}).
    If we determine $ |\widetilde{C}_1|^2$, $|\widetilde{C}_2|^2 $, $Re[\widetilde{C}_1\widetilde{C}_2^*]$, and $Im[\widetilde{C}_1\widetilde{C}_2^*] $, respectively, then, using their $t$ dependence, we can derive the time independent coefficients.
    The result is summarized as follows:
\begin{align} \begin{split}
    |\widetilde{C}_1|^2 &\Rightarrow  |C_1|^2, \left|\frac{q}{p}\right|^2|\bar C_1|^2, Re[C_1 \frac{q^*}{p^*}\bar C_1^*],Im[C_1 \frac{q^*}{p^*}\bar C_1^*]
\\  |\widetilde{C}_2|^2 &\Rightarrow  |C_2|^2, \left|\frac{q}{p}\right|^2|\bar C_2|^2, Re[C_2 \frac{q^*}{p^*}\bar C_2^*],Im[C_2 \frac{q^*}{p^*}\bar C_2^*]
\\  Re[\widetilde{C}_1\widetilde{C}_2^*] &\Rightarrow  Re[C_1 C_2^*], \left|\frac{q}{p}\right|^2 Re[\bar C_1 \bar C_2^*],
\\& Re[C_1 \frac{q^*}{p^*} \bar C_2^*]+Re[\frac{q}{p} \bar C_1 C_2^*],
    Im[C_1 \frac{q^*}{p^*} \bar C_2^*]-Im[\frac{q}{p} \bar C_1 C_2^*]
\\  Im[\widetilde{C}_1\widetilde{C}_2^*] &\Rightarrow  Im[C_1 C_2^*], \left|\frac{q}{p}\right|^2 Im[\bar C_1 \bar C_2^*],
\\& Re[C_1 \frac{q^*}{p^*} \bar C_2^*]-Re[\frac{q}{p} \bar C_1 C_2^*], 
    Im[C_1 \frac{q^*}{p^*} \bar C_2^*]+Im[\frac{q}{p} \bar C_1 C_2^*].
\end{split} \end{align}
    To determine these quantities, we need a huge number of statistics.

    In Examples 1-5 (and $\bar B^0 \to \Lambda_c^+ \bar p$ in Example 6), we integrated over the time dependence, and took sum over $B^0$ ($\bar B^0$) decay events and their CP conjugate.
    However, if we take difference between the $B^0$ decay events and their CP conjugate instead of sum, we obtain
\begin{align} \begin{split}
 &  \frac{X_B}{1+X_B^2} Im[\frac{q}{p} C_1\bar C_1^*],  \hspace{6.5em}
    \frac{X_B}{1+X_B^2} Im[\frac{q}{p} C_2\bar C_2^*], \\
 &   \frac{1}{1+X_B^2} (Re[C_1 C_2^*]-Re[\bar C_1 \bar C_2^*])+ \frac{X_B}{1+X_B^2} (Im[\frac{p}{q}C_1 \bar C_2^* ]-Im[\frac{p^*}{q^*}\bar C_1  C_2^* ]), \\
 &   \frac{1}{1+X_B^2} (Im[C_1 C_2^*]-Im[\bar C_1 \bar C_2^*])- \frac{X_B}{1+X_B^2} (Re[\frac{p}{q}C_1 \bar C_2^* ]-Re[\frac{p^*}{q^*}\bar C_1  C_2^* ]),
\end{split} \end{align}
    where $X_B=\Delta m_B/\Gamma_B$.
    If we detect at least one of them, it means that CP is violated.

    In the Examples 1-5, we showed some simple processes.
    We can determine the parameters more precisely by using not only these processes but also $\tau^+\to\pi^+\pi^0\bar\nu_\tau$ and other processes.



    In Section \ref{only one}, we studied the case that only one of two fermions is unstable.
    $B\to\tau \mu$ mode is new physics itself.
    So if anything, $D_4$ value is very significant for understanding it.
    
    In Section \ref{baryon}, we studied the baryon modes.
    These modes contain the non-perturbative QCD effects to pollute the possible new physics effect.
    If we suppose that there is no new physics, we can test the factorization and the pole model, for example.
    On the other hand, if the experimental result highly differs from these predictions, we should consider the new physics effect.
    Recently, the lattice gauge theory predicts some $B$ decay processes \cite{lattice}.
    We hope that the lattice gauge theory predicts precisely the $B$ meson baryonic decays in near future. 
    If so, we can search for new physics, precisely.


    We emphasize that, to discover new physics, it is necessary to determine as many physical quantities as possible, and compare them to the SM predictions.
    Moreover, it is preferable to be done by the unified form for simplicity, facility, and practicality.
    This paper will help this process.


\appendix

\section{differential branching ratios for $\bar f_a$ and $f_b$ decays } \label{App:f-decay}
The differential Branching ratio for the process $\bar f_a \to a+\mathrm{anything}$ and $ f_b \to b+\mathrm{anything}$ in $\bar f_a$ and $f_b$ rest frame, respectively, are \cite{sanda-2};
\begin{align} \begin{split}  \label{fa}
\frac{dBr(\bar f_a \to a+\mathrm{anything})}{d^3k_{a}}  
&\equiv Br_a\frac{2}{\pi m_a^3 \lambda_a}\left[ G_1^a(y_a)+  \mathbf{s}^a \cdot \hat {\mathbf{k}}_a  G_2^a(y_a) \right], 
\\ 
\frac{dBr( f_b \to b+\mathrm{anything})}{d^3k_{b}}  
&\equiv Br_b\frac{2}{\pi m_b^3 \lambda_b}\left[ G_1^b(y_b)-  \mathbf{s}^b \cdot \hat {\mathbf{k}}_b  G_2^b(y_b) \right], 
\end{split} \end{align}
where $G_1^{a,b}(y_{a,b})$ and $G_2^{a,b}(y_{a,b})$ are the functions of $y_{a,b}=2E_{a,b}/m_{a,b}$, $\lambda_{a,b}$ are defined as 
\begin{eqnarray}
\lambda_{a,b}=\int dy_{a,b} y_{a,b}^2 G_1^{a,b}(y_{a,b}),
\end{eqnarray}
$\hat {\mathbf{k}}_{a,b}=  \mathbf{k}_{a,b}  /|\mathbf{k}_{a,b}|$ where $k_{a,b}$ are the momentum of the particle $a$ and $b$, respectively.

We note here that physical vector quantities which we treat in this process are only $ \mathbf{s}^{a,b} $ and $ \hat {\mathbf{k}}_{a,b} $.
The only scalar made by these vector quantities is $ \mathbf{s}^{a,b} \cdot \hat {\mathbf{k}}_{a,b} $
So, we can explain the differential branching ratio, Eq. (\ref{fa}) by only two terms which are proportional to $G_1^{a,b}(y_{a,b})$ and $  \mathbf{s}^{a,b} \cdot \hat {\mathbf{k}}_{a,b}  G_2^a(y_{a,b})$, respectively.

\end{document}